\documentclass[10 pt, twocolumn]{IEEETran}

\def \be {\begin{equation}}
\def \ee {\end{equation}}

\def \nn {\nonumber}

\usepackage{ifpdf}

\usepackage[dvips]{graphicx}
\usepackage{color}

\usepackage[cmex10]{amsmath}
\usepackage{amssymb}

\begin{document}
%
\title{\LARGE \bf A Differential-Cascaded Paradigm for Control of Nonlinear Systems}
%
%
%

\author{Hanlei~Wang, Wei Ren, and Chien Chern Cheah
\thanks{H. Wang is with the Science and Technology on Space Intelligent Control Laboratory,
Beijing Institute of Control Engineering,
Beijing 100094, China (e-mail: hlwang.bice@gmail.com).}
\thanks{W. Ren is with the Department of Electrical and Computer Engineering,
University of California at Riverside, Riverside, CA 92521 USA (e-mail: ren@ece.ucr.edu).}
\thanks{C. C. Cheah is with the School of Electrical and Electronic Engineering,
Nanyang Technological University, Singapore 639798 (e-mail:
ecccheah@ntu.edu.sg).}
}
\maketitle

\begin{abstract}
This paper focuses on developing a new paradigm motivated by investigating the consensus problem of networked Lagrangian systems with time-varying delay and switching topologies. We present adaptive controllers with piecewise continuous or arbitrary times differentiable control torques for realizing consensus of Lagrangian systems, extending the results in the literature. This specific study motivates the formulation of a new paradigm referred to as forwardstepping, which is shown to be a systematic tool for solving various nonlinear control problems. One distinctive point associated with forwardstepping is that the order of the reference dynamics is typically specified to be equal to or higher than that of the original nonlinear system, and the reference dynamics and the nonlinear system are governed by a differential/dynamic-cascaded structure. The order invariance or increment of the specified reference dynamics with respect to the nonlinear system and their differential/dynamic-cascaded structure expands significantly the design freedom and thus facilitates the seeking of solutions to many nonlinear control problems which would otherwise often be intractable.
\end{abstract}


\section{Introduction}

The investigation of consensus of networked systems over the past years has yielded a substantial number of interesting results (see, e.g., \cite{Olfati-Saber2004_TAC,Moreau2005_TAC,Ren2005_TAC,Lin2005_TAC,Chopra2006,Tian2008_TAC,Lestas2010_AUT,Munz2011b_TAC,Nuno2011_TAC,Yu2010_AUT,Wang2013b_AUT,Abdessameud2014_TAC,Wang2014_TAC,Nuno2018_TCST}).
Two important and also challenging issues in consensus of multiple dynamical systems concern the handling of time delay (which is typically encountered in either the communication channel or information processing) and the accommodation of switching topologies. Numerous results have been presented for resolving the consensus of linear systems with communication delay, input delay, and/or switching topologies (see, e.g., \cite{Olfati-Saber2004_TAC,Tian2008_TAC,Lestas2010_AUT,Munz2011b_TAC}). The consideration of uncertain Lagrangian systems with constant or time-varying communication delay and/or switching topology occurs in, e.g., \cite{Nuno2011_TAC,Abdessameud2014_TAC,Wang2014_TAC,Wang2020b_AUT}. The intertwining between the issue of communication delay and switching topology and that of the dynamics of each system generally renders the consensus problem more challenging (see, e.g., \cite{Nuno2011_TAC,Wang2020_AUT}); this becomes more prominent in the case of time-varying delay, either in the aspect of controller design or in that of analysis (e.g., \cite{Abdessameud2014_TAC,Wang2020b_AUT,Wang2020_AUT}). In particular, time-varying delay is typically unknown and might involve discontinuity, and similarly, switching topology also introduces discontinuous quantities. In the case of switching topology, a topology-independent
solution is proposed in \cite{Wang2017_CAC}, and for handling
both the time-varying delay and switching
topology, delay/topology-independent solutions are provided in \cite{Wang2020b_AUT,Wang2020_AUT}.

In this paper, we start by investigating and extending the results in \cite{Wang2017_CAC,Wang2020b_AUT} with differentiable reference velocities in the case of piecewise uniformly continuous time-varying delay and switching topology. We develop delay/topology-independent adaptive controllers with arbitrary times differentiable reference velocities (consequently, piecewise continuous or arbitrary times differentiable control torques) for consensus of multiple Lagrangian systems with time-varying delay and switching topologies. These adaptive controllers rely upon the definition of the reference acceleration and velocity by a dynamic system that is of arbitrary order. This is for generating control torques that can be piecewise continuous or arbitrary times differentiable in the case of discontinuous time-varying delay and switching topologies. The objective is achieved by extending the differentiable reference velocity defined in \cite{Wang2017_CAC,Wang2020b_AUT} to be arbitrary times differentiable (by a procedure referred to as adding differentiators). The distinctive point of our result, in comparison with \cite{Abdessameud2014_TAC,Nuno2018_TCST}, lies in its delay-independent property and the consideration of switching topologies. In the context involving only switching topologies, a specific solution is presented in \cite{Abdessameud2018b_TAC}, which extends the first-order solution in \cite{Wang2017_CAC} to yield a reference velocity that is arbitrarily times differentiable using an arbitrary-order dynamic system without involving relative velocity measurement, by combining the basic framework in \cite{Wang2017_CAC} with the results for linear integrator systems (see, e.g., \cite{Liu2017_TCybernetics,Abdessameud2018_TAC}). Our starting example concerns both the time-varying delay and switching topologies, and we unify the design/analysis for these two important issues under a general framework. For instance, a systematic analysis approach is introduced to formalize the stability analysis of differential/dynamic-cascaded systems with arbitrary degree, and the exploitation of the Laplacian structure of the resultant consensus dynamics establishes the connection with the results for single-integrator, double-integrator, and high-order integrator systems (e.g., \cite{Munz2011b_TAC,Liu2017_TCybernetics,Abdessameud2018_TAC}). 

At this stage, we might wonder why this practice is possible. Taking the first-order solutions in \cite{Wang2017_CAC,Wang2020b_AUT} as an example, the challenges for handling time-varying delay and switching topology are essentially the same in the sense that both of them do not favor the differential operation. For instance, consider a function with unknown time-varying delay $x(t-T(t))$ with $T(t)$ being the delay and its derivative can be written as ${d}/{dt}[x(t-T(t))]=[1-\dot T(t)]\dot x(t-T(t))$,
which involves the derivative of the delay. In the case of switching topologies, the differential operation of the switching action becomes infinite at the switching points and thus not realizable.
  On the other hand, both the unknown delay and switching topology are compatible well with the integral operation. Hence, the second purpose of this study is to seek the fundamental nature of this line of results (e.g., \cite{Wang2017_CAC,Wang2020b_AUT,Abdessameud2018b_TAC}), and we formally propose a new paradigm that is referred to as forwardstepping. From the perspective of design, this paradigm is opposite to the standard backstepping design \cite{Krstic1995_Book,Kokotovic2001_AUT}, and the resultant feedback involves one integral operation in its most basic or lowest-order form such as \cite{Wang2017_CAC,Wang2020b_AUT}, due to the fact that a differentiator is added for designing the feedback. In the general case, the feedback relies on the integral of a dynamic system, in contrast to the dependence on the differential operation in the backstepping design. From the perspective of analysis, we perform a procedure of reducing the differential degree concerning the states of lower subsystems involved in the upper ones, and we refer to this procedure as degree reduction or adding integral operators.

Forwardstepping is characterized by defining reference dynamics with the order being equal to or greater than that of the original dynamical system and by the differential-cascaded architecture or structure that connects the reference dynamics and the system dynamics, and hence the control input typically involves the integral operation (and algebraic operation) of the reference dynamics. The choice of the reference dynamics [which would converge to typical target dynamics (see, e.g., \cite{Astolfi2003_TAC}) as the reference converge to the involved state], due to the versatility of the differential-cascaded or dynamic-cascaded structure and the order invariance or increment permitted by the differential-cascaded structure, is significantly enlarged in comparison with the traditional order reduction confined by the cascaded structure (for instance, backstepping design), and this facilitates the seeking of solutions to many control problems that would otherwise often be intractable. In a general context, the reference dynamics can be specified based upon those results for linear systems, as is typically done in the traditional context of order reduction; the distinctive point due to the dynamic-cascaded structure is that the results concerning high-order linear systems can be employed for designing reference dynamics for low-order nonlinear systems. In the control of many physical systems such as robot manipulators and spacecraft, the controlled natural/physical dynamics (e.g., the controlled dynamics of robot manipulators under the standard proportional-derivative control in \cite{Takegaki1981_ASME}) can often be a promising start for seeking the appropriate reference dynamics, as will be demonstrated in our result (see also \cite{Astolfi2003_TAC} in the context of order reduction). The analysis associated with forwardstepping typically involves that of a dynamic-cascaded system with nonzero degree, and the proposed degree reduction can be applied to complete the analysis. Forwardstepping, due to its strong connection with the integral action and the design freedom permitted by the choice of reference dynamics with order invariance or increment, is believed to have the potential of becoming a systematic tool for control of nonlinear systems involving, e.g., discontinuities, time-varying uncertainty, unavailable state measurement, and certain kind of underactuation. In addition, the order invariance or increment in specifying reference dynamics and the differential-cascaded architecture yields a new perspective beyond the conventional one relying on the state-space representation. In particular, we demonstrate the application of forwardstepping to control (consensus, tracking, and distributed tracking) of a class of nonlinear systems without involving zero dynamics (which can be written as the normal form without involving zero dynamics; see, e.g., \cite[Chapter~6]{Slotine1991_Book}), namely Lagrangian systems, thrust-propelled vehicles, and spacecraft; this yields solutions to several open problems, e.g., global attitude tracking of a single spacecraft actuated by reaction wheels without angular velocity measurement and distributed tracking of uncertain Lagrangian systems on directed topologies using physical measurement only. A preliminary version of our result was presented in \cite{Wang2019_ACC}, and our present paper expands it to further detail the forwardstepping approach with additional examples, and to address the tracking control of mechanical systems without velocity measurement and distributed tracking of networked Lagrangian systems.

\section{Preliminaries}

\subsection{Graph Theory}

We first briefly introduce the graph theory \cite{Olfati-Saber2004_TAC,Ren2005_TAC,Ren2008_Book} in the context of $n$ dynamical systems. As is now standard, we employ a directed graph $\mathcal G=(\mathcal V,\mathcal E)$ to describe the interaction topology among the $n$ systems with $\mathcal V=\{1,\dots,n\}$ being the vertex set that denotes the collection of the $n$ systems and $\mathcal E\subseteq \mathcal V\times \mathcal V$ the edge set that denotes the information interaction among the $n$ systems. Denote by $\mathcal N_i=\{j|(i,j)\in\mathcal E\}$ the set of neighbors of the $i$-th system. A graph is said to contain a directed spanning tree if there exists a vertex $k^\ast\in\mathcal V$ such that any other vertex of the graph has a directed path to $k^\ast$. The weighted adjacency matrix $\mathcal W=[w_{ij}]$ associated with the graph $\mathcal G$ is defined in accordance with the rule that $w_{ij}>0$ if $j\in\mathcal N_i$, and $w_{ij}=0$ otherwise. In addition, we adopt the typical assumption concerning the diagonal entries of $\mathcal W$, namely $w_{ii}=0$, $\forall i=1,\dots,n$. Upon the definition of $\mathcal W$, the Laplacian matrix $\mathcal L_w=[ \ell_{w,ij}]$ associated with the graph $\mathcal G$ is defined as $\ell_{w,ij}=\Sigma_{k=1}^n w_{ik}$ if $i=j$, and $\ell_{w,ij}=-w_{ij}$ otherwise.
In the case of switching topology, the interaction graph among the $n$ systems is time-varying. Denote by ${\mathcal G}_S=\{\mathcal G_1,\dots,\mathcal G_{n_s}\}$ the set of the interaction graphs among the $n$ systems, and these graphs share the same vertex set $\mathcal V$ but their edge sets may be different. The union of a collection of graphs $G_{i_1}, \dots,G_{i_s}$ with $i_s\le n_s$ is a graph with the vertex set being given by $\mathcal V$ and edge set given by the union of the edge sets of $G_{i_1}, \dots,G_{i_s}$.  Denote by $t_0,t_1,t_2,\dots$ with $t_0=0$ an infinite sequence of times at which the interaction graph switches, and suppose that this sequence satisfies the standard dwell-time condition that $t_{\kappa+1}-t_\kappa \ge T_D$, $\forall \kappa=0,1,\dots$ with $T_D$ being certain positive constant (see \cite{Ren2008_Book} for the detail).

\subsection{Equations of Motion of Lagrangian Systems}

The equations of motion of the $i$-th Lagrangian system can be written as \cite{Slotine1991_Book,Spong2006_Book}
\be
\label{eq:1}
M_i(q_i)\ddot q_i+C_i(q_i,\dot q_i)\dot q_i+g_i(q_i)=\tau_i
\ee
where $q_i\in\mathcal R^m$ is the generalized position (or configuration), $M_i(q_i)\in \mathcal R^{m\times m}$ is the inertia matrix, $C_i(q_i,\dot q_i)\in \mathcal R^{m\times m}$ is the Coriolis and centrifugal matrix, $g_i(q_i)\in \mathcal R^m$ is the gravitational torque, and $\tau_i\in \mathcal R^m$ is the exerted control torque. Three well-known properties associated with the dynamics (\ref{eq:1}) are listed as follows (see, e.g., \cite{Slotine1991_Book,Spong2006_Book}).

\emph{Property 1:} The inertia matrix $M_i(q_i)$ is symmetric and uniformly positive definite.

\emph{Property 2:} The Coriolis and centrifugal matrix $C_i(q_i,\dot q_i)$ can be suitably chosen so that the matrix $\dot M_i(q_i)-2C_i(q_i,\dot q_i)$ is skew-symmetric.

\emph{Property 3:} The dynamics (\ref{eq:1}) depend linearly on an unknown constant dynamic parameter vector $\vartheta_i$, which yields
\be
M_i(q_i)\dot\zeta+C_i(q_i,\dot q_i)\zeta+g_i(q_i)=Y_i(q_i,\dot q_i,\zeta,\dot \zeta)\vartheta_i
\ee
where $\zeta\in\mathcal R^m$ is a differentiable vector, $\dot \zeta$ is the derivative of $\zeta$, and $Y_i(q_i,\dot q_i,\zeta,\dot \zeta)$ is the regressor matrix.

\section{A Motivating Example: Consensus of Networked Lagrangian Systems}

We start with the study of consensus of networked Lagrangian systems with time-varying delay and switching topology, and the essential idea is connected with specifying reference dynamics with the order being equal to or greater than that of Lagrangian systems, which illustrates the properties of a new paradigm referred to as forwardstepping, differing from the standard backstepping design.

\subsection{Adaptive Control With Differentiable Reference Velocity}

We consider the example of consensus of $n$ Lagrangian systems with switching topology and time-varying delay. The delays are assumed to be piecewise uniformly continuous and uniformly bounded. The backstepping design (as illustrated and formulated in \cite{Kokotovic2001_AUT,Krstic1995_Book}) views the Lagrangian dynamics as a cascaded system (this is also equivalent to the standard state-space representation of Lagrangian systems\footnote{The typical state-space representation of Lagrangian systems, with $x_i=[q_i^T,\dot q_i^T]^T$ as the state, can be written as
\be
\underbrace{\frac{d}{dt}\begin{bmatrix}q_i\\
\dot q_i\end{bmatrix}=\begin{bmatrix}\dot q_i\\
M_i^{-1}(q_i)[-C_i(q_i,\dot q_i)\dot q_i-g_i(q_i)]
\end{bmatrix}+\begin{bmatrix}0_m\\ M_i^{-1}(q_i)\end{bmatrix}\tau_i}_{\dot x_i=f_i(x_i)+B_i(x_i)\tau_i}.
\ee})
\be
\label{eq:aa2}
\begin{cases}
\dot q_i=\dot q_i\\
M_i(q_i)\ddot q_i+C_i(q_i,\dot q_i)\dot q_i+g_i(q_i)=\tau_i
\end{cases}
\ee
while the forwardstepping views the Lagrangian dynamics as a differential/dynamic-cascaded system and one form among such systems can be written as
\be
\label{eq:aa3}
\begin{cases}
\ddot q_i=\ddot q_i\\
M_i(q_i)\ddot q_i+C_i(q_i,\dot q_i)\dot q_i+g_i(q_i)=\tau_i.
\end{cases}
\ee

The backstepping design starts with taking $\dot q_i$ in the first subsystem of (\ref{eq:aa2}) as the virtual control with the desired virtual control specified as
\be
\dot q_{r,i}={-\Sigma_{j\in\mathcal N_i(t)}w_{ij}(t)[q_i-q_j(t-T_{ij})]}
\ee
where $T_{ij}$ is the time-varying delay from the $j$-th system to the $i$-th system that may be discontinuous, and this yields
\be
\dot q_i=-\Sigma_{j\in\mathcal N_i(t)}w_{ij}(t)[q_i-q_j(t-T_{ij})]+(\dot q_i-\dot q_{r,i}).
\ee
The second step of backstepping design is to stabilize the second subsystem of (\ref{eq:aa2}) with respect to the error $\dot q_i-\dot q_{r,i}$. Consider the typical passivity-based design for the second subsystem (namely applying the well-known control in \cite{Slotine1987_IJRR}), and then the differentiation of the desired virtual control $\dot q_{r,i}$ is involved. The torque control can be given as \cite{Slotine1987_IJRR}
\be
\label{eq:aa6}
\begin{cases}
\tau_i=-K_i(\dot q_i-\dot q_{r,i})+Y_i(q_i,\dot q_i,\dot q_{r,i},\ddot q_{r,i})\hat\vartheta_i\\
\dot{\hat\vartheta}_i=-\Gamma_i Y_i^T (q_i,\dot q_i,\dot q_{r,i},\ddot q_{r,i})(\dot q_i-\dot q_{r,i})
\end{cases}
\ee
where $K_i$ and $\Gamma_i$ are symmetric positive definite matrices, and $\hat\vartheta_i$ is the estimate of $\vartheta_i$.
 Note that the derivative of the desired virtual control, namely $\ddot q_{r,i}$ is required in (\ref{eq:aa6}) and it becomes infinite as the topology is switching or the delay abruptly changes, and in addition, the differentiation of $q_j(t-T_{ij})$, i.e., $d/dt[q_j(t-T_{ij})]=(1-\dot T_{ij})\dot q_j(t-T_{ij})$ involves the derivative of the time-varying delay which is difficult to be measured/calculated. The closed-loop dynamics yielded by backstepping design are a cascaded system
\be
\begin{cases}
{\dot q_i=-\Sigma_{j\in\mathcal N_i(t)}w_{ij}(t)[q_i-q_j(t-T_{ij})]}+s_i\\
M_i(q_i)\dot s_i+C_i(q_i,\dot q_i)s_i=-K_is_i+Y_i(q_i,\dot q_i,\dot q_{r,i},\ddot q_{r,i})\Delta\vartheta_i\\
\dot{\hat\vartheta}_i=-\Gamma_i Y_i^T (q_i,\dot q_i,\dot q_{r,i},\ddot q_{r,i})s_i
\end{cases}
\ee
with $s_i=\dot q_i-\dot q_{r,i}$ and $\Delta\vartheta_i=\hat\vartheta_i-\vartheta_i$.

The first step of forwardstepping, differing from that of backstepping, takes $\ddot q_i$ of the first subsystem of (\ref{eq:aa3}) as the virtual control; in addition, the considered dynamics in the first step, namely $\ddot q_i=\ddot q_i$ is of second order with respect to $q_i$, in contrast to the first step of backstepping (which concerns first-order dynamics $\dot q_i=\dot q_i$). The desired virtual control $\dot z_i$ for the second-order dynamics $\ddot q_i=\ddot q_i$ is defined by also a second-order dynamic system
\be
\label{eq:aa10}
\dot z_i=-\alpha \dot q_i-\Sigma_{j\in\mathcal N_i(t)}w_{ij}(t)[\xi_i-\xi_j(t-T_{ij})]
\ee
where $\xi_i$ is defined as \cite{Chopra2006}
\be
\label{eq:4}
\xi_i=\dot q_i+\alpha q_i
\ee
with $\alpha$ being a positive design constant, which yields
\be
\ddot q_i=-\alpha \dot q_i-\Sigma_{j\in\mathcal N_i(t)}w_{ij}(t)[\xi_i-\xi_j(t-T_{ij})]+\dot s_i
\ee
where $s_i=\dot q_i-z_i$. In the second step of forwardstepping, we stabilize the second subsystem of (\ref{eq:aa3}) with respect to $\ddot q_i-\dot z_i$ or its integral $s_i=\dot q_i-z_i$ with the control being specified as \cite{Slotine1987_IJRR} with new reference velocity and acceleration
\be
\begin{cases}
\tau_i=-K_is_i+Y_i(q_i,\dot q_i,z_i,\dot z_i)\hat\vartheta_i\\
\dot{\hat\vartheta}_i=-\Gamma_i Y_i^T (q_i,\dot q_i,z_i,\dot z_i)s_i.
\end{cases}
\ee
Note that the derivation of the control does not involve differentiation of discontinuous quantities since the derivation of $\dot z_i$ and $z_i$ rely on (\ref{eq:aa10}) and the integral operation of (\ref{eq:aa10}), respectively, in contrast to backstepping (which involves the differentiation of the discontinuous desired virtual control $\dot q_{r,i}$). The closed-loop dynamics can be written as
\be
\label{eq:aa12}
\begin{cases}
{\dot \xi_i=-\Sigma_{j\in\mathcal N_i(t)}w_{ij}(t)[\xi_i-\xi_j(t-T_{ij})]}+\dot s_i\\
M_i(q_i)\dot s_i+C_i(q_i,\dot q_i)s_i=-K_is_i+Y_i(q_i,\dot q_i,z_i,\dot z_i)\Delta\vartheta_i\\
\dot{\hat\vartheta}_i=-\Gamma_i Y_i^T (q_i,\dot q_i,z_i,\dot z_i)s_i.
\end{cases}
\ee
The above system is referred to as differential-cascaded or dynamic-cascaded since the interconnection component $\dot s_i$ in the first subsystem involves the derivative of the states of the lower two subsystems (namely $s_i$ and $\hat\vartheta_i$), due to which the analysis of (\ref{eq:aa12}) typically requires degree reduction concerning $\dot s_i$ in the first subsystem. By applying forwardstepping, the differentiation of discontinuous quantities (thus possibly yielding infinite control input) and the involvement of measurement of the derivative of the unknown time-varying delay are avoided or suitably accommodated (see \cite{Wang2020_AUT} for the technical details), in contrast to the application of backstepping.

\subsection{Adaptive Control With Twice Differentiable Reference Velocity}

In this subsection, we design adaptive controllers with twice differentiable reference velocity (namely the vector $z_i$ defined below), extending the previous result (in Sec. III-A and \cite{Wang2020_AUT}). This is achieved by specifying the desired virtual control $\ddot z_i$ via third-order reference dynamics (greater than the order of Lagrangian systems) for the third-order dynamics
\be
\label{eq:aa13}
\dddot q_i=\dddot q_i.
\ee

We start by considering the case of fixed topology with time-varying delay. Define a twice differentiable vector $z_i$ by a dynamic system [which is of third order with respect to $q_i$, and acts as reference dynamics or defines the desired virtual control input for (\ref{eq:aa13})]
\begin{align}
\label{eq:3}
\ddot z_i=&-(\alpha+\beta) \ddot q_i-\alpha\beta \dot q_i\nn\\
&-\Sigma_{j\in\mathcal N_i}w_{ij}[\dot \xi_i+\beta \xi_i-\beta\xi_j(t-T_{ij})]
\end{align}
where $\alpha$ and $\beta$ are positive design constants. Define
\be
\label{eq:5}
s_i=\dot q_i-z_i.
\ee
The adaptive controller is given as
\begin{align}
\label{eq:6}
\tau_i=&-K_i s_i+Y_i(q_i,\dot q_i,z_i,\dot z_i)\hat \vartheta_i\\
\label{eq:7}
\dot{\hat \vartheta}_i=&-\Gamma_i Y_i^T(q_i,\dot q_i,z_i,\dot z_i)s_i.
\end{align}
 The basic adaptive structure of the proposed controller is the same as that in \cite{Slotine1987_IJRR}, and the distinctive point lies in the reference velocity and acceleration (i.e., $z_i$ and $\dot z_i$) defined by (\ref{eq:3}), which is developed by adding differentiators based on the results for consensus of linear systems (e.g., \cite{Liu2017_TCybernetics}). This potentially enhances the connection between nonlinear systems and linear consensus control.

The dynamics of the $i$-th system can be described by
\be
\label{eq:9}
\begin{cases}
\ddot \xi_i=-\beta\dot \xi_i-\Sigma_{j\in\mathcal N_i}w_{ij}[\dot \xi_i+\beta\xi_i-\beta \xi_j(t-T_{ij})]+\ddot s_i\\
M_i(q_i)\dot s_i+C_i(q_i,\dot q_i)s_i=-K_i s_i+Y_i(q_i,\dot q_i,z_i,\dot z_i)\Delta \vartheta_i\\
\dot {\hat \vartheta}_i=-\Gamma_i Y_i^T(q_i,\dot q_i,z_i,\dot z_i)s_i.
\end{cases}
\ee
The system (\ref{eq:9}) is referred to as a dynamic-cascade or differential-cascade system (in the sense of \cite{Wang2020b_AUT,Wang2020_AUT}) with degree two and is distinguished from the typical cascade systems in that the interconnection component $\ddot s_i$ involves the second derivative of the state of the lower two subsystems. For the convenience of stability analysis, we transform the above dynamic-cascade system with degree two to a dynamic-cascade system with degree one as
\be
\label{eq:11}
\begin{cases}
\dot \xi_i= y_i+\dot s_i\\
\dot  y_i=-\beta y_i-\Sigma_{j\in\mathcal N_i}w_{ij}[ y_i+\beta \xi_i-\beta \xi_j(t-T_{ij})]\\
\quad \quad { } -(\beta+\Sigma_{j\in{\mathcal N}_i}w_{ij})\dot s_i\\
M_i(q_i)\dot s_i+C_i(q_i,\dot q_i)s_i
=-K_i s_i+Y_i(q_i,\dot q_i,z_i,\dot z_i)\Delta \vartheta_i\\
\dot {\hat \vartheta}_i=-\Gamma_i Y_i^T(q_i,\dot q_i,z_i,\dot z_i)s_i
\end{cases}
\ee
where $y_i=\dot \xi_i-\dot s_i=\dot z_i+\alpha\dot q_i$.

\emph{Remark 1:} For generating differentiable control torques even if the time-varying delay involves discontinuity, we introduce a dynamic system with its order higher than the one in \cite{Wang2020b_AUT,Wang2017_CAC} and this design leads to the dynamic-cascaded system (\ref{eq:9}). To facilitate the stability analysis, we perform a degree reduction transformation so that the dynamic-cascade system with degree two is transformed to a dynamic-cascade system with degree one. The comparison between (\ref{eq:9}) and (\ref{eq:11}) shows that as the interconnection component $\ddot s_i$ is transferred along the upward direction, it is equivalent to integral operations and the consequence is that $\ddot s_i$ is changed to $\dot s_i$ at an upper position of the system. It may seem interesting that our design and analysis involves two opposite operations: adding differentiators in design, and adding integrators in analysis.

\emph{Remark 2:} The adaptive controller given by (\ref{eq:6}) and (\ref{eq:7}) relies on the calculation of $z_i$ and $\dot z_i$ based on (\ref{eq:3}) using pure integrals. The calculation of $\dot z_i$ does not require acceleration measurement and the measurement of $q_i$ and $\dot q_i$ is adequate for implementation (by a pure integral operation). 


\emph{Theorem 1:} Suppose that the interaction graph among the $n$ Lagrangian systems contains a directed spanning tree. Then, the adaptive controller given by (\ref{eq:6}) and (\ref{eq:7}) with $z_i$ being given by (\ref{eq:3}) guarantees the consensus of the $n$ Lagrangian systems, i.e., $q_i-q_j\to 0$ and $\dot q_i\to 0$ as $t\to\infty$, $\forall i,j=1,\dots,n$.

\emph{Proof:} In accordance with the typical practice (see, e.g., \cite{Ortega1989_Aut,Slotine1987_IJRR}), we consider the Lyapunov-like function candidate $V_i=(1/2)s_i^T M_i(q_i)s_i+(1/2)\Delta\vartheta_i^T\Gamma_i^{-1}\Delta \vartheta_i$ and its derivative along the trajectories of the second and third subsystems of (\ref{eq:9}) can be written as (using Property 2) $\dot V_i=-s_i^T K_is_i\le 0$, which yields the result that $s_i\in{\mathcal L}_2\cap{\mathcal L}_\infty$ and $\hat\vartheta_i\in{\mathcal L}_\infty$, $\forall i$. From the upper two subsystems in (\ref{eq:11}), we obtain the following dynamic system
\be
\label{eq:12}
\begin{cases}
\dot \xi= y+\dot s\\
\dot y=-\beta y- \Psi_D(y,\xi)-[(\beta I_n+\mathcal D_w)\otimes I_m]\dot s
\end{cases}
\ee
where $I_n$ and $I_m$ are the $n\times n$ and $m\times m$ identity matrices, respectively, $\otimes$ denotes the standard Kronecker product, $\xi=[\xi_1^T,\dots,\xi_n^T]^T$, $y=[y_1^T,\dots,y_n^T]^T$, $s=[s_1^T, \dots, s_n^T]^T$,
$
\Psi_D(y,\xi)
={\rm column}\{\Sigma_{j\in\mathcal N_i}w_{ij}[ y_i+\beta \xi_i-\beta \xi_j(t-T_{ij})], i=1,\dots,n\}
$,
and $\mathcal D_w={\rm diag}[\Sigma_{j\in\mathcal N_i}w_{ij}, i=1,\dots,n]$ is the degree matrix \cite{Ren2008_Book}.
Consider the system (\ref{eq:12}) with $\dot s=0$
\be
\label{eq:14}
\begin{cases}
\dot \xi= y\\
\dot  y=-\beta y- \Psi_D(y,\xi).
\end{cases}
\ee
 Following the procedure in \cite{Liu2017_TCybernetics}, we obtain that
\be
\label{eq:17}
\begin{cases}
\beta\dot\xi=-\beta[\beta\xi-(\dot \xi+\beta \xi)]\\
\frac{d}{dt}(\dot \xi+\beta \xi)=-\Psi_D(\dot \xi,\xi).
\end{cases}
\ee
The formulation of (\ref{eq:17}) is for exhibiting a Laplacian structure that is essentially the same as  \cite{Liu2017_TCybernetics}; in particular, in the case of no communication delay, the coefficient matrix of the linear system (\ref{eq:17}) can be written as
$
A=-\begin{bmatrix}\beta I_n & -\beta I_n
\\-\mathcal W & \mathcal D_{w}\end{bmatrix}\otimes I_m.
$
In accordance with \cite{Munz2011b_TAC}, the Laplacian structure of (\ref{eq:17}) yields the robustness with respect to arbitrary bounded time-varying delay (i.e., piecewise uniformly continuous and uniformly bounded delay), and the result that $\xi_i-\xi_j$, $\dot \xi_i$, and $\ddot \xi_i$ uniformly asymptotically converge to zero as $t\to\infty$, $\forall i,j$; hence the system (\ref{eq:17}) or (\ref{eq:14}) is a uniformly marginally stable system of the first kind in the sense that the state uniformly converges to a constant vector (see \cite{Wang2020_AUT}). {For the system (\ref{eq:12}) with $[\xi^T,y^T]^T$ as the state and $[\dot s^T,[(\beta I_n+\mathcal D_w)\otimes I_m]\dot s^T]^T$ as the input, using \cite[Proposition~3]{Wang2020_AUT}}, we obtain that $y\in \mathcal L_\infty$ and $-\beta y-\Psi_D(y,\xi)\in{\mathcal L_\infty}$ since $\int_0^t \dot s(\sigma)d\sigma=s-s(0)\in \mathcal L_\infty$. This in turn implies that $y_i\in{\mathcal L}_\infty$ and $\Sigma_{j\in\mathcal N_i} w_{ij}[\xi_i-\xi_j(t-T_{ij})]\in \mathcal L_\infty$, $\forall i$. We rewrite (\ref{eq:3}) as
\begin{align}
\label{eq:19}
\ddot z_i=&-(\alpha+\beta) \dot z_i-\alpha\beta z_i-\Sigma_{j\in\mathcal N_i}w_{ij}[y_i+\beta \xi_i\nn\\
&-\beta\xi_j(t-T_{ij})]-\alpha\beta s_i-(\alpha+\beta +\Sigma_{j\in\mathcal N_i}w_{ij})\dot s_i
\end{align}
with $z_i$ and $\dot z_i$ as the output, and we note that the system
$
\ddot z_i=-(\alpha+\beta)\dot z_i-\alpha\beta z_i
$
is a standard exponentially stable LTI (linear time-invariant) system. In accordance with the standard input-output properties of exponentially stable and strictly proper linear systems \cite[p.~59]{Desoer1975_Book}, we obtain that the first portion of the output of (\ref{eq:19}) corresponding to the input $-\Sigma_{j\in\mathcal N_i}w_{ij}[y_i+\beta \xi_i-\beta\xi_j(t-T_{ij})]-\alpha\beta s_i$ (bounded) is bounded. {Using \cite[Proposition~1]{Wang2020_AUT}}, we obtain that the second portion of the output of (\ref{eq:19}) corresponding to the integral-bounded input $-(\alpha+\beta +\Sigma_{j\in\mathcal N_i}w_{ij})\dot s_i$ [due to the fact that $\int_0^t \dot s_i(\sigma)d\sigma=s_i-s_i(0)\in\mathcal L_\infty$]
is also bounded. By the standard superposition principle for linear systems, we obtain that $z_i\in{\mathcal L_\infty}$ and $\dot z_i\in{\mathcal L}_\infty$, $\forall i$. Hence, $\dot q_i=z_i+s_i\in{\mathcal L}_\infty$, $\forall i$. {Using Property 1 and the standard property associated with the considered Lagrangian systems that $C_i(q_i,\dot q_i)$ and $g_i(q_i)$ are bounded independent of $q_i$, we obtain from the third subsystem of (\ref{eq:11})} that $\dot s_i\in\mathcal L_\infty$ and thus $\ddot q_i=\dot z_i+\dot s_i\in\mathcal L_\infty$, implying that $s_i$ and $\dot q_i$ are uniformly continuous, $\forall i$. Using the properties of square-integrable and uniformly continuous functions \cite[p.~232]{Desoer1975_Book}, we obtain that $s_i\to 0$ as $t\to\infty$, $\forall  i$. From (\ref{eq:19}), we obtain that $\ddot z_i\in\mathcal L_\infty$, and from (\ref{eq:7}), we obtain that $\dot {\hat \vartheta}_i\in{\mathcal L}_\infty$, $\forall i$. It can thus be shown that $q_i$, $\dot q_i$, $z_i $, $\dot z_i$, and $\hat\vartheta_i$ are uniformly continuous, $\forall i$. From the third subsystem of (\ref{eq:11}), it can be shown that $\dot s_i$ is uniformly continuous by using Property 1, and hence $\ddot q_i=\dot z_i+\dot s_i$ is also uniformly continuous, $\forall i$. From (\ref{eq:7}), we know that $\dot{\hat \vartheta}_i$ is uniformly continuous, $\forall i$. By differentiating the third subsystem of (\ref{eq:11}) and using Property 1, we obtain that $\ddot s_i\in{\mathcal L}_\infty$ and $\ddot s_i$ is piecewise uniformly continuous due to its dependence on $\ddot z_i$ which is piecewise uniformly continuous according to (\ref{eq:3}), $\forall i$. Using Barbalat's lemma \cite{Slotine1991_Book}, we obtain that $\dot s_i\to 0$ as $t\to\infty$, $\forall i$. For the system (\ref{eq:12}), we obtain that $\dot \xi_i\to 0$ and $\dot y_i\to 0$ as $t\to\infty$ using {\cite[Proposition~3]{Wang2020_AUT}}, $\forall i$, and hence, $y_i=\dot \xi_i-\dot s_i\to 0$ and $\Sigma_{j\in\mathcal N_i}w_{ij}\beta [\xi_i-\xi_j(t-T_{ij})]\to 0$ as $t\to\infty$, $\forall i$. Differentiating (\ref{eq:4}) with respect to time yields
$
\ddot q_i+\alpha \dot q_i=\dot \xi_i
$ and using the input-output properties of exponentially stable and strictly proper linear systems \cite[p.~59]{Desoer1975_Book}, we obtain that $\dot q_i\to 0$ as $t\to\infty$, $\forall i$, and it can also be shown by the standard practice that $q_j-q_j(t-T_{ij})=\int_0^{T_{ij}}\dot q_j(t-\sigma)d\sigma\to 0$ as $t\to\infty$, $\forall j\in\mathcal N_i$, $\forall i$. Therefore, we have that $\Sigma_{j\in\mathcal N_i}w_{ij}\beta [\xi_i-\xi_j(t-T_{ij})]\to \Sigma_{j\in\mathcal N_i}w_{ij}\alpha\beta (q_i-q_j)\to 0$ as $t\to\infty$, $\forall i$. This implies that $\alpha\beta(\mathcal L_w\otimes I_m) q\to 0$ as $t\to\infty$ with $q=[q_1^T,\dots,q_n^T]^T$, and using the standard properties of $\mathcal L_w$ (see, e.g., \cite{Ren2005_TAC,Ren2008_Book}), we obtain that $q_i-q_j\to 0$ as $t\to\infty$, $\forall i,j$. \hfill {\small $\blacksquare$}

\emph{Remark 3:} The definition (\ref{eq:3}) can further be reshaped as
\begin{align}
\label{eq:23}
\ddot z_i=&-(\alpha+\beta) \ddot q_i-\alpha\beta \dot q_i-\Sigma_{j\in\mathcal N_i}w_{ij}[\ddot q_i+(\alpha+\beta)\dot q_i\nn\\
&+\alpha\beta q_i-\alpha\beta q_j(t-T_{ij})].
\end{align}
This is simply the fact that the velocities of the neighbors can be removed, yielding the consequence that the relative damping $\Sigma_{j\in\mathcal N_i}w_{ij}\beta (\dot q_i-\dot q_j)$ becomes the absolute damping $\Sigma_{j\in\mathcal N_i}w_{ij}\beta \dot q_i$.  The practice (\ref{eq:23}) is mainly due to the result for double-integrator systems in \cite{Liu2017_TCybernetics} as well as its extension to high-order integrator systems (see, e.g., \cite{Abdessameud2018_TAC}), which will be further exploited later to develop a solution with arbitrary order. In our result above, damping is considered to be rather general (i.e., generalized damping) and the derivative of a quantity, as is known, can act as damping with respect to this quantity; for instance, $\dot \xi_i$ acts as the role of damping with respect to the quantity $\xi_i$, facilitating the extension of \cite{Liu2017_TCybernetics}.

For accommodating both the time-varying delay and switching topology, we redefine $z_i$ as
\begin{align}
\label{eq:a1}
\ddot z_i=&-(\alpha+\beta) \dot z_i-\alpha\beta \dot q_i\nn\\
&-\Sigma_{j\in\mathcal N_i(t)}w_{ij}(t)[\dot z_i+\alpha \dot q_i+\beta \xi_i-\beta\xi_j(t-T_{ij})].
\end{align}
{The comparison between (\ref{eq:a1}) and (\ref{eq:3}) shows that the acceleration $\ddot q_i$ is replaced {with} $\dot z_i$ in (\ref{eq:a1}), which is for handling the time-varying nature of $w_{ij}(t)$ in the case of switching topologies}. With the adaptive controller given by (\ref{eq:6}) and (\ref{eq:7}), the dynamics of the $i$-th system can then be described by
\be
\label{eq:a2}
\begin{cases}
\dot \xi_i= y_i+\dot s_i\\
\dot  y_i=-\beta y_i-\Sigma_{j\in\mathcal N_i(t)}w_{ij}(t)\\
\quad \quad{ }\times[ y_i+\beta \xi_i-\beta \xi_j(t-T_{ij})] +\alpha\dot s_i\\
M_i(q_i)\dot s_i+C_i(q_i,\dot q_i)s_i\\
=-K_i s_i+Y_i(q_i,\dot q_i,z_i,\dot z_i)\Delta \vartheta_i\\
\dot {\hat \vartheta}_i=-\Gamma_i Y_i^T(q_i,\dot q_i,z_i,\dot z_i)s_i.
\end{cases}
\ee

\emph{Theorem 2:} Suppose that there exist an infinite number of uniformly bounded intervals $[t_{\kappa_h},t_{\kappa_{h+1}})$, $h=1,2,\dots$ with $t_{\kappa_1}=t_0$ satisfying the property that the union of the interaction graphs in each interval contains a directed spanning tree. Then, the adaptive controller given by (\ref{eq:6}) and (\ref{eq:7}) with $z_i$ being given by (\ref{eq:a1}) guarantees the consensus of the $n$ Lagrangian systems, i.e., $q_i-q_j\to 0$ and $\dot q_i\to 0$ as $t\to\infty$, $\forall i,j=1,\dots,n$.

The proof of Theorem 2 mainly relies on the analysis of the following dynamic system
\be
\label{eq:a3}
\begin{cases}
\dot \xi= y+\dot s\\
\dot y=-\beta y- \Psi_D^\ast(y,\xi)+\alpha\dot s
\end{cases}
\ee
where $\Psi_D^\ast(y,\xi)={\rm column}\{\Sigma_{j\in\mathcal N_i(t)}w_{ij}(t)[ y_i+\beta \xi_i-\beta \xi_j(t-T_{ij})],i=1,\dots,n\}$, and the coefficient of $\dot s$ holds constant due to the definition (\ref{eq:a1}) even if the topology is switching. This is distinguished from the case in (\ref{eq:12}) that involves the degree matrix of the graph. With $\dot s=0$, we obtain a system that is similar to (\ref{eq:17})
\be
\label{eq:a4}
\begin{cases}
\beta\dot\xi=-\beta[\beta\xi-(\dot \xi+\beta \xi)]\\
\frac{d}{dt}(\dot \xi+\beta \xi)=-\Psi_D^\ast(\dot \xi,\xi).
\end{cases}
\ee
Exploiting the Laplacian structure of (\ref{eq:a4}) and using the results in \cite{Munz2011b_TAC,Wang2020_AUT,Liu2017_TCybernetics}, the remaining proof can be performed by following similar procedures as in the proof of Theorem 1.

\subsection{Adaptive Control With Arbitrary Times Differentiable Reference Velocity}

We now present arbitrary-order solutions to the considered problem so that the reference velocity is arbitrary times differentiable (i.e., the arbitrary-order derivatives of the reference velocity exist). Generally, in comparison with the low-order solution, the high-order solution yields smoother control torques at the expense of relatively slower dynamic response. In addition, as is shown later, the high-order solution facilitates the handling of underactuation of certain nonlinear systems and more complicated control problems such as distributed tracking. The most attractive feature of the high-order solution, in our opinion, may lie in the design freedom that it provides (e.g., for accommodating the case of high-order nonlinear systems), beyond the smoother control input as emphasized in \cite{Abdessameud2018b_TAC}.

 For generating $\ell$ times differentiable reference velocities, we specify $\ell+1$-th-order reference dynamics for the $\ell+1$-th-order dynamics \be
 d^{\ell+1} q_i/dt^{\ell+1}=d^{\ell+1} q_i/dt^{\ell+1},\ell=1,2,\dots.\ee
 In particular, by adding $\ell$ differentiators, we introduce a dynamic system with both the relative position and velocity measurement as
\begin{align}
\label{eq:a5}
&\frac{d^\ell z_i}{d t^\ell}=-\alpha_{i,\ell-1} \frac{d^{\ell-1} z_i}{d t^{\ell-1}}-\dots-\alpha_{i,1}  \frac{d z_i}{dt}-\alpha_{i,0} \dot q_i\nn\\
&-\Sigma_{j\in\mathcal N_i(t)}w_{ij}(t)[\frac{d^{\ell-1} z_i}{d t^{\ell-1}}+\alpha_{i,\ell-1}\frac{d^{\ell-2} z_i}{d t^{\ell-2}}+\dots+\alpha_{i,2}\frac{d z_i}{dt}\nn\\
&+\alpha_{i,1}\dot q_i+\alpha_{i,0} q_i- (\alpha_{i,0}/\kappa_{i,0})\dot q_j(t-T_{ij})-\alpha_{i,0}q_j(t-T_{ij})]
\end{align}
or a dynamic system without relative velocity measurement as
\begin{align}
\label{eq:24}
&\frac{d^\ell z_i}{d t^\ell}=-\alpha_{i,\ell-1} \frac{d^{\ell-1} z_i}{d t^{\ell-1}}-\dots-\alpha_{i,1}  \frac{dz_i}{dt}-\alpha_{i,0} \dot q_i\nn\\
&-\Sigma_{j\in\mathcal N_i(t)}w_{ij}(t)[\frac{d^{\ell-1} z_i}{d t^{\ell-1}}+\alpha_{i,\ell-1}\frac{d^{\ell-2} z_i}{d t^{\ell-2}}+\dots+\alpha_{i,2}\frac{d z_i}{dt}\nn\\
&+\alpha_{i,1}\dot q_i+\alpha_{i,0} q_i- \alpha_{i,0}q_j(t-T_{ij})]
\end{align}
where $\alpha _{i,r}$, $r=0,1,\dots,\ell-1$ are positive design constants being chosen such that the polynomial
\be
\label{eq:25}
\theta^{\ast\ell}+\alpha_{i,\ell-1}\theta^{\ast\ell-1}+\dots+\alpha_{i,1} \theta^\ast+\alpha_{i,0}
\ee
with $1,\alpha_{i,\ell-1},\dots,\alpha_{i,0}$ being the coefficients and $\theta^\ast$ the free variable is a Hurwitz polynomial with all the roots being real, which are denoted by $-\kappa_{i,r}$, $r=0,1,\dots,\ell-1$.

\emph{Theorem 3:}  Suppose that there exist an infinite number of uniformly bounded intervals $[t_{\kappa_h},t_{\kappa_{h+1}})$, $h=1,2,\dots$ with $t_{\kappa_1}=t_0$ satisfying the property that the union of the interaction graphs in each interval contains a directed spanning tree. Then, the adaptive controller given by (\ref{eq:6}) and (\ref{eq:7}) with $z_i$ being given by (\ref{eq:a5}) or (\ref{eq:24}) guarantees the consensus of the $n$ Lagrangian systems, i.e., $q_i-q_j\to 0$ and $\dot q_i\to 0$ as $t\to\infty$, $\forall i,j=1,\dots,n$.

\emph{Proof:} Following the same procedures as in the proof of Theorem 1, we have that $s_i\in{\mathcal L}_2\cap \mathcal L_\infty$ and $\hat\vartheta_i\in{\mathcal L}_\infty$, $\forall i$. We first consider the case that $z_i$ and $\dot z_i$ are given by (\ref{eq:a5}). The dynamics of the $i$-th system become [more general than (\ref{eq:9})]
\be
\label{eq:26}
\begin{cases}
\frac{d^{\ell+1} q_i}{d t^{\ell+1}}=-\alpha_{i,\ell-1} \frac{d^{\ell-1} z_i}{d t^{\ell-1}}-\dots-\alpha_{i,1}  \frac{d z_i}{dt}-\alpha_{i,0} \dot q_i\\
-\Sigma_{j\in\mathcal N_i(t)}w_{ij}(t)[\frac{d^{\ell-1} z_i}{d t^{\ell-1}}+\alpha_{i,\ell-1}\frac{d^{\ell-2} z_i}{d t^{\ell-2}}+\dots+\alpha_{i,2}\frac{dz_i}{dt}+\alpha_{i,1}\dot q_i\\
+\alpha_{i,0} q_i- (\alpha_{i,0}/\kappa_{i,0})\dot q_j(t-T_{ij})-\alpha_{i,0}q_j(t-T_{ij})]+\frac{d^\ell s_i}{dt^\ell}\\
M_i(q_i)\dot s_i+C_i(q_i,\dot q_i)s_i=-K_i s_i+Y_i(q_i,\dot q_i,z_i,\dot z_i)\Delta \vartheta_i\\
\dot {\hat \vartheta}_i=-\Gamma_i Y_i^T(q_i,\dot q_i,z_i,\dot z_i)s_i
\end{cases}
\ee
which is a differential/dynamic-cascaded system with degree $\ell$. With {degree reduction} to the first subsystem of (\ref{eq:26}) (i.e., adding $\ell-1$ integral operators to $\frac{d^{\ell+1} q_i}{dt^{\ell+1}}-\frac{d^\ell s_i}{dt^\ell}$), we obtain
\begin{align}
\label{eq:27}
\begin{cases}
\dot q_i=\dot q_i\\
\ddot q_i=\frac{d z_i}{dt}+\dot s_i\\
\frac{d^2 z_i}{dt^2}=\frac{d^2 z_i}{dt^2}\\
\quad\vdots\\
\frac{d^{\ell} z_i}{d t^{\ell}}=-\alpha_{i,\ell-1} \frac{d^{\ell-1} z_i}{d t^{\ell-1}}-\dots-\alpha_{i,1}  \frac{d z_i}{dt}-\alpha_{i,0} \dot q_i\\
-\Sigma_{j\in\mathcal N_i(t)}w_{ij}(t)[\frac{d^{\ell-1} z_i}{d t^{\ell-1}}+\alpha_{i,\ell-1}\frac{d^{\ell-2} z_i}{d t^{\ell-2}}+\dots+\alpha_{i,2}\frac{dz_i}{dt}\\
+\alpha_{i,1}\dot q_i+\alpha_{i,0} q_i- (\alpha_{i,0}/\kappa_{i,0})\dot q_j(t-T_{ij})-\alpha_{i,0}q_j(t-T_{ij})],
\end{cases}
\end{align}
and with $\dot s_i=0$, we have from (\ref{eq:27}) that
\begin{align}
\label{eq:28}
&\frac{d}{dt}\bigg[\frac{d^\ell q_i}{d t^\ell}+\alpha_{i,\ell-1}\frac{d^{\ell-1} q_i}{d t^{\ell-1}}+\dots+\alpha_{i,0} q_i\bigg]\nn\\
&=-\Sigma_{j\in\mathcal N_i(t)}w_{ij}(t)\bigg[\frac{d^\ell q_i}{d t^\ell}+\alpha_{i,\ell-1}\frac{d^{\ell-1} q_i}{d t^{\ell-1}}+\cdots\nn\\
&+\alpha_{i,0} q_i- (\alpha_{i,0}/\kappa_{i,0})\dot q_j(t-T_{ij})-\alpha_{i,0}q_j(t-T_{ij})\bigg].
\end{align}
By the assumption concerning (\ref{eq:25}), the polynomial (\ref{eq:25}) can be written as the standard form
$
(\theta^\ast+\kappa_{i,\ell-1})\cdots (\theta^\ast+\kappa_{i,0}).
$
Following \cite{Liu2017_TCybernetics,Abdessameud2018_TAC}, we obtain a dynamic system as [based on (\ref{eq:28})]
\begin{align}
\label{eq:a6}
\begin{cases}
(\kappa_{i,1}\cdots \kappa_{i,\ell-1})\frac{d}{dt}(\dot q_i+\kappa_{i,0} q_i)\\
=-\kappa_{i,1}[(\kappa_{i,1}\cdots \kappa_{i,\ell-1})(\dot q_i+\kappa_{i,0} q_i)\\
\quad-(\kappa_{i,2}\cdots \kappa_{i,\ell-1})(p+\kappa_{i,1})(\dot q_i+\kappa_{i,0} q_i)]\\
\quad\vdots\\
\kappa_{i,\ell-1}\frac{d}{dt}[(p+\kappa_{i,\ell-2})\cdots(p+\kappa_{i,1})(\dot q_i+\kappa_{i,0}q_i)]\\
=-k_{i,\ell-1}[\kappa_{i,\ell-1}(p+\kappa_{i,\ell-2})\cdots(p+\kappa_{i,1})(\dot q_i+\kappa_{i,0}q_i)\\
\quad-(p+\kappa_{i,\ell-1})\cdots(p+\kappa_{i,1})(\dot q_i+\kappa_{i,0}q_i)]\\
\frac{d}{dt}\big[(p+\kappa_{i,\ell-1})\cdots(p+\kappa_{i,1})(\dot q_i+\kappa_{i,0}q_i)\big]\\
=-\Sigma_{j\in\mathcal N_i(t)}w_{ij}(t)\big[(p+\kappa_{i,\ell-1})\cdots(p+\kappa_{i,1})(\dot q_i+\kappa_{i,0}q_i)\\
\quad-(\alpha_{i,0}/\kappa_{i,0})\dot q_j(t-T_{ij})-\alpha_{i,0} q_j(t-T_{ij})\big]
\end{cases}
\end{align}
where $p$ denotes the standard differential operator. The system (\ref{eq:a6}), similar to (\ref{eq:17}) and (\ref{eq:a4}), exhibits a Laplacian structure with respect to the vectors $\psi_i=[(\kappa_{i,1}\cdots \kappa_{i,\ell-1})(\dot q_i+\kappa_{i,0}q_i)^T,\dots,\kappa_{i,\ell-1}(p+\kappa_{i,\ell-2})\cdots(p+\kappa_{i,1})(\dot q_i+\kappa_{i,0}q_i)^T,(p+\kappa_{i,\ell-1})\cdots(p+\kappa_{i,1})(\dot q_i+\kappa_{i,0}q_i)^T]^T$, $i=1,\dots,n$.
Then in accordance with the result in \cite{Munz2011b_TAC} and standard linear system theory, we obtain from the system (\ref{eq:a6}) that $q_i-q_j\to 0$ and $\frac{d^r q_i}{d t^r}\to 0$, $r=1,\dots,\ell+1$ as $t\to\infty$, $\forall i,j$. In the case that $z_i$ and $\dot z_i$ are generated by (\ref{eq:24}), we can establish a dynamic system similar to (\ref{eq:a6}) as
\begin{align}
\label{eq:30}
\begin{cases}
(\kappa_{i,0}\cdots \kappa_{i,\ell-1})\dot q_i\\
=-\kappa_{i,0}[(\kappa_{i,0}\cdots \kappa_{i,\ell-1})q_i\\
\quad-(\kappa_{i,1}\cdots \kappa_{i,\ell-1})(p+\kappa_{i,0})q_i]\\
\quad\vdots\\
\kappa_{i,\ell-1}\frac{d}{dt}[(p+\kappa_{i,\ell-2})\cdots(p+\kappa_{i,0})q_i]\\
=-k_{i,\ell-1}[\kappa_{i,\ell-1}(p+\kappa_{i,\ell-2})\cdots(p+\kappa_{i,0})q_i\\
\quad-(p+\kappa_{i,\ell-1})\cdots(p+\kappa_{i,0})q_i]\\
\frac{d}{dt}\big[(p+\kappa_{i,\ell-1})\cdots(p+\kappa_{i,0})q_i\big]\\
=-\Sigma_{j\in\mathcal N_i(t)}w_{ij}(t)\big[(p+\kappa_{i,\ell-1})\cdots(p+\kappa_{i,0})q_i\\
\quad-\alpha_{i,0} q_j(t-T_{ij})\big].
\end{cases}
\end{align}
The result similar to the first case then follows. The subsequent proof can be completed by following similar procedures as in the proofs of Theorem 1 and Theorem 2. \hfill{\small $\blacksquare$}


\emph{Remark 4:} As shown in \cite{Wang2020b_AUT}, manipulability of a dynamic system becomes an important issue if it is expected to interact with an external subject (e.g., a human operator); in particular the infinite manipulability with degree one is a basic requirement for ensuring easy and safe human-system interaction (e.g., bilateral teleoperation). The analysis of manipulability is based on input-output gains of dynamical systems, and in this sense, the dynamic-cascade system with high degree can serve as a basis for manipulability analysis.  In particular, we can modify the definition 
(\ref{eq:3}) as
\begin{align}
\label{eq:31}
\ddot z_i=&-(\alpha+\beta) \ddot q_i-\alpha\beta \dot q_i-\Sigma_{j\in\mathcal N_i}w_{ij}[\dot \xi_i+\beta \xi_i\nn\\
&-\beta\xi_j(t-T_{ij})]+\lambda_{\mathcal M}(\dot q_i-z_i)
\end{align}
where $\lambda_{\mathcal M}$ is a positive number that acts as the weighting factor for balancing between the consensus of the $n$ Lagrangian systems and intensity of the infinite manipulability. The definition (\ref{eq:a1}), (\ref{eq:a5}), or (\ref{eq:24}) can be similarly reformulated. The basic idea is to incorporate a square-integrable function that might be integral unbounded; see \cite{Wang2020b_AUT} for the detail.

\emph{Remark 5:} For multiple Lagrangian systems with switching topologies, a recent result is presented in \cite{Abdessameud2018b_TAC}, which extends the differentiable reference velocity in \cite{Wang2017_CAC} to the one that is arbitrary times differentiable without involving relative velocity measurement. Our result considers the time-varying delay in addition to switching topologies, with or without relative velocity measurement, and more importantly our result uncovers the fundamental nature of and motivates a general paradigm behind this line of results (e.g., \cite{Wang2017_CAC,Abdessameud2018b_TAC,Wang2020b_AUT}).

\section{Forwardstepping Paradigm}

\subsection{Forwardstepping and Its Connection With Control of Nonlinear Systems}

The basic idea of the methodology, as demonstrated in \cite{Wang2017_CAC,Wang2020_AUT}, is to avoid the differential operation in specifying the control input, leading to the integral operation (and algebraic operation). We might recall that the solution of a differential equation can be considered as a generalized integral operation. Intuitively, the integral operation of a variable or function, as compared with its counterpart---differential operation, shows prominent advantages as involving discontinuity, delay uncertainty, and challenging measurement concerning a variable or function (e.g.,  measurement of the derivative of a variable tends to be problematic). From the perspective of analysis, we observe an opposite operation and in fact we need to analyze the stability of a differential/dynamic-cascade system with nonzero degree (with the interconnection component involving the derivative or high-order derivatives of the states), for which we propose degree reduction analysis (i.e., adding integral operators in analysis).

The basic mechanism behind motivates the formulation of a new systematic design/analysis paradigm referred to as forwardstepping so as to fully exploit its (potential) applications in other contexts. The new paradigm can be considered as the counterpart of the standard backstepping paradigm (see, e.g., \cite{Krstic1995_Book}). The design procedure of the forwardstepping approach
mainly involves adding differentiators as compared with the standard backstepping approach that basically relies on adding integrators. The forwardstepping approach also differs from the forwarding approach (see, e.g., \cite{Teel1992_IFAC,Teel1996_TAC,Sepulchre1997_Book,Kaliora2004_TAC}) in that the basic design of the forwarding approach is opposite to that of the backstepping approach yet does not rely on adding differentiators for deriving the feedback (e.g., forwarding  is referred to as integrator forwarding in \cite{Sepulchre1996_IFAC}). As a consequence of the new design procedure, the closed-loop dynamics are typically dynamic-cascaded, and due to the proposed rigorous analysis tools (e.g., degree reduction analysis) the resultant controllers can often be ensured to be stable/convergent with milder conditions (benefiting from the dynamic-cascade perspective), as shown in Sec. III and the examples in the sequel. 

Consider a general $k$-th order nonlinear system without involving zero dynamics given as (see, e.g., \cite[Chapter~6]{Slotine1991_Book})
\be
\label{eq:33}
\frac{d^k x}{dt^k}=\chi(x^\ast)+ \psi(x^\ast)u
\ee
with $x\in\mathcal R$, $u\in\mathcal R$, $x^\ast=[x,dx/dt,\dots,d^{k-1}x/dt^{k-1}]^T$, and $\chi(\cdot)$ and $\psi(\cdot)$ are nonlinear mappings. The forwardstepping approach views the above system as a dynamic-cascaded or differential-cascaded system (adding $\ell$ differentiators with respect to the state $d^{k-1} x/dt^{k-1}$, $\ell=1,2,\dots$)
\begin{align}
\frac{d^{k+\ell-1} x}{dt^{k+\ell-1}}=&\frac{d^{k+\ell-1} x}{dt^{k+\ell-1}}\nn\\
\quad\vdots\nn\\
\frac{d^k x}{dt^k}=&\frac{d^k x}{dt^k}\nn\\
\frac{d^k x}{dt^k}=&\chi(x^\ast)+\psi(x^\ast)u,
\end{align}
and the design is conducted with $\frac{d^{k+\ell-1} x}{dt^{k+\ell-1}},\dots,\frac{d^k x}{dt^k}$ (or part of these quantities) as the virtual control. As a comparison, we recall the backstepping approach which typically views the above system as a cascaded system or as the normal form (i.e., adding $k-1$ integrators; see \cite[p.~249]{Slotine1991_Book})
\begin{align}
\frac{d x}{dt}=&\frac{d x}{dt}\nn\\
\quad\vdots\nn\\
\frac{d^{k-1} x}{dt^{k-1}}=&\frac{d^{k-1} x}{dt^{k-1}}\nn\\
\frac{d^k x}{dt^k}=&\chi(x^\ast)+\psi(x^\ast)u
\end{align}
with $\frac{d x}{dt},\dots,\frac{d^{k-1} x}{dt^{k-1}}$ as the virtual control.

The above different perspectives lead to different design and analysis for forwardstepping and backstepping. The rationale behind the generation of possibly new solutions using forwardstepping is connected with the integrability of functions (whether or not the integral of functions can be explicitly expressed using elementary functions); forwardstepping yields reference vectors (e.g., $\dot z_i$ in the previous results) that are not necessarily integrable while the application of backstepping typically yields reference vectors with integrable forms.

On the other hand, forwardstepping may be combined with backstepping (and also other available design approaches) to yield new solutions to many important control problems. For instance, their combination can be used to solve the problem of consensus of parametric-pure/strict-feedback nonlinear systems with switching topology and time-varying delay; the arbitrary times differentiable reference (acting as the desired virtual control) can be obtained using the proposed arbitrary-order reference dynamics by applying forwardstepping, and the tracking with respect to the reference is ensured by the adaptive backstepping control (see, e.g., \cite{Kanellakopoulos1991_TAC,Kokotovic2001_AUT}); the stability analysis can be conducted with degree reduction.

The benefit using forwardstepping is associated with that of using the integral operation concerning time-varying gains and functions; as compared with the differential operation adopted by backstepping, using the integral operation does not involve the derivatives of the time-varying gains (which can be either unknown or infinite) or the derivatives of functions/variables [e.g., the derivative of the position---velocity is often difficult to be measured; if a variable contains unknown time-varying delay, e.g., $x(t-T(t))$, its derivative would involve the derivative of the unknown delay]. The details can be witnessed in the examples of the paper.

\subsection{A General Framework for the Application of Forwardstepping}

In the application of forwardstepping, the design concerning the reference dynamics (e.g., $z_i$-dynamics) plays an important role, and the design freedom associated with the reference dynamics achieves separation between the applied control action and stability analysis of the closed-loop system. The closed-loop system is typically a differential/dynamic-cascaded one with the order of the reference dynamics being equal to or higher than that of the system dynamics while the nonlinearity of the reference dynamics is generally weaker than that of the system dynamics. This differs from the standard cascade-system perspective that relies on order reduction in the sense that in that case, the order of the reference dynamics is typically lower than that of the system dynamics. For instance, in the standard Slotine and Li adaptive controller \cite{Slotine1987_IJRR}, the reference velocity dynamics are of first order, which is lower than that of the robot dynamics (second order), and consequently the differential operation of the reference velocity dynamics is required for deriving the reference acceleration. Similar cases occur in the backstepping approach \cite{Krstic1995_Book} and the immersion and invariance (I\&I) approach \cite{Astolfi2003_TAC} (with the dimension of the specified target dynamics strictly lower than that of the original nonlinear system), typically yielding a standard cascade closed-loop system.

The differential/dynamic-cascaded structure associated with forwardstepping allows order invariance or increment of reference dynamics with respect to the original nonlinear dynamical system and also the versatility of choice of reference dynamics. In a general context, we can specify the reference dynamics based on the results for linear systems, as is typically done, yet with the order being equal to or higher than that of the nonlinear system benefiting from the dynamic-cascaded structure. In control of many physical systems (e.g., robot manipulators or spacecraft), the choice of reference dynamics can be started with those results for controlled physical systems (the closed-loop dynamics are, in some sense, still physical, and one typical example is the controlled robot manipulator with the implementation of proportional-derivative control \cite{Takegaki1981_ASME}; see also \cite{Astolfi2003_TAC} for some detailed discussions), and often the nonlinearity reduction concerning the controlled physical systems is involved in determining the appropriate reference dynamics. Another potentially promising direction may be the application of forwardstepping in connection with the existing nonlinear design approaches; for instance, the application of forwardstepping in a context with the I\&I approach would enlarge the freedom of choice of reference dynamics, whose order can be higher than that of those considered in \cite{Astolfi2003_TAC}.

If we confine our attention to the static consensus problem (i.e., position consensus with its derivative and high-order derivatives converging to zero), we can routinely consider the following design procedure for general dynamical systems:
\begin{enumerate}

\item Introduce the $z_i$-dynamics by adding $\ell$ differentiators ($\ell=1,2,\dots$) and also specify the desired consensus dynamics (linear or nonlinear consensus protocol from the currently available literature) at the same level as the $z_i$-dynamics where the order of the desired consensus dynamics can be equal to or higher than that of the dynamical systems\footnote{{For instance, the specified consensus dynamics in (\ref{eq:9}) yield a third-order system with respect to $q_i$ while the dynamics of Lagrangian systems are of second order with respect to $q_i$, $i=1,\dots,n$.}};

\item Apply {the passivity-based or other nonlinear design approaches} for generating the control input with the necessary feedback concerning the quantification of the error between the specified state and $z_i$ [for instance, the variable $s_i$, its high-order derivatives (for instance, $\dot s_i$), or its filtered form];

\item Perform the stability analysis of the resultant dynamic-cascade system using the degree reduction analysis.

\end{enumerate}
This design procedure yields a control input that is at least $\ell-1$ times differentiable and a closed-loop system that is dynamic-cascaded with degree $\ell$.

The applications of forwardstepping to other problems can be conducted in a similar way. For the nonlinear system given as (\ref{eq:33}), in the first step, we can consider the differential-cascaded system
\begin{align}
\frac{d^{k+\ell-1}x}{dt^{k+\ell-1}}=&\frac{d^{k+\ell-1}x}{dt^{k+\ell-1}}\nn\\
\frac{d^k x}{dt^k}=&\chi(x^\ast)+\psi(x^\ast)u.
\end{align}
and design reference dynamics concerning a vector $z$ with order invariance or increment, namely $\frac{d^\ell z}{dt^\ell}=\mathcal Z(z,\dots, \frac{d^{\ell-1} z}{d t^{\ell-1}},x,\dots,\frac{d^{k+l-2} x}{d t^{k+l-2}})$ for $\frac{d^{k+\ell-1}x}{dt^{k+\ell-1}}$ such that
\begin{align}
\label{eq:43}
\frac{d^{k+\ell-1}x}{dt^{k+\ell-1}}=&\mathcal Z(z,\dots, \frac{d^{\ell-1}z}{d t^{\ell-1}},x,\dots,\frac{d^{k+\ell-2} x}{d t^{k+\ell-2}})\nn\\
&+\mathcal F\left(\frac{d^{\ell}(\frac{d^{k-1}x}{dt^{k-1}}-z)}{dt^{\ell}},\dots,\frac{d^{k-1}x}{dt^{k-1}}-z\right)\nn\\
\frac{d}{dt}\left(\frac{d^{k-1}x}{dt^{k-1}}-z\right)=&\chi(x^\ast)-\dot z+\psi(x^\ast)u
\end{align}
where $\mathcal Z(z,\dots, \frac{d^{\ell-1} z}{d t^{\ell-1}},x,\dots,\frac{d^{k+l-2} x}{d t^{k+l-2}})$ governs the desired dynamics associated with the control objective, and $\mathcal F$ is a mapping governing the interconnection component or the differential/dynamic-cascaded structure. In the second step, we choose the passivity-based design or other nonlinear design approaches to stabilize the second subsystem of (\ref{eq:43}) concerning $\frac{d^{k-1}x}{d t^{k-1}}-z$. Finally, we perform the analysis of the resultant dynamic-cascaded closed-loop system with degree $\ell$ using degree reduction.

\emph{Remark 6:}  The high-order derivatives of $x$ with the order higher than $\frac{d^{k-1}x}{d t^{k-1}}$ involved can either be avoided during the integral operation of the $z$-dynamics (since only $z$ and $\dot z$ are typically involved as stabilizing the second subsystem) or be replaced with the derivative or high-order derivatives of $z$ (which are available since they are constructed).

 In the next section, we demonstrate the application of forwardstepping to adaptive tracking of a point mass without velocity measurement, consensus of thrust-propelled vehicles (TPVs), model-based tracking control of two categories of mechanical systems without velocity measurement, and distributed tracking of networked Lagrangian systems. These different applications are unified and systematically handled in the sense of the specification of the $z_i$-dynamics or $z$-dynamics with the order equal to or higher than that of the original dynamical system and of the resultant differential/dynamic-cascaded closed-loop dynamics; in particular, the $z_i$-dynamics or $z$-dynamics (namely reference dynamics) are specified to be either linear or nonlinear with order invariance or increment, depending on the specific contexts.

\section{Applications of Forwardstepping}

\subsection{Adaptive Tracking of a Point Mass Without Velocity Measurement}

Consider a point mass governed by
\be
\label{eq:a7}
m\ddot x=u
\ee
where $x\in {\mathcal R^{m}}$ denotes the position, $m\in {\mathcal R}$ the mass, and $u\in {\mathcal R}$ the control input. We consider the adaptive position tracking problem with $x_d\in {\mathcal R}$ denoting the desired position, and assume that $x_d$, $\dot x_d$, $\ddot x_d$, and $\dddot x_d$ are all bounded. Consider the following control law
\be
\label{eq:a8}
u=\hat m \dot z-k y
\ee
where $k$ is a positive design constant, $\hat m$ is the estimate of $m$, $z\in\mathcal R$ is defined by the following third-order dynamic system
\be
\label{eq:a9}
\ddot z=\dddot x_d-\alpha_2(\dot z-\ddot x_d)-\alpha_1(z-\dot x_d)-\alpha_0(x-x_d),
\ee
with $\alpha_0$, $\alpha_1$, and $\alpha_2$ being positive design constants which are chosen such that $\theta^{\ast 3}+\alpha_2\theta^{\ast 2}+\alpha_1\theta^\ast+\alpha_0$ with $\theta^\ast$ the free variable is a Hurwitz polynomial, and $y\in \mathcal R$ is given by the standard passive filter as
\be
\dot y+\lambda_f y=\lambda _f (\dot x-z)
\ee
where $\lambda_f$ is a positive design constant.
The adaptation law for $\hat m$ is given as
\be
\dot{\hat m}=-\gamma^\ast \dot z s
\ee
where $\gamma^\ast$ is a positive design constant and $s=\dot x-z$.

\emph{Remark 7:} The adaptive output feedback control of linear systems with relative degree two is a standard problem in the literature and some solutions to this problem are given in, e.g., \cite{Ioannou1996_Book,Slotine1991_Book}. {Our solution is derived under the guidance of forwardstepping and by exploiting the design freedom provided by the third-order dynamic system (\ref{eq:a9}) (different from the standard model reference approach in \cite{Ioannou1996_Book,Slotine1991_Book}), and the proposed solution does not require the nonlinear high-gain feedback in \cite[p.~359]{Ioannou1996_Book} or overparametrization in the model-reference framework \cite{Ioannou1996_Book,Slotine1991_Book}}. The second term in (\ref{eq:a8}) follows the standard approach in \cite{Berghuis1993_SCL}, and in addition, the calculation of $\hat m$ does not need the velocity measurement since by the standard integral by parts
$
\hat m(t)
=\hat m (0)-\gamma^\ast ( z x- z z)|_0^t+\gamma^\ast\int_0^t x(\sigma) \ddot z(\sigma)d\sigma
$
and we note that $\ddot z$ does not involve velocity measurement. {In addition, we can solve the adaptive consensus problem for  double-integrator systems without involving velocity measurement by combining the result here and that in Sec. III}

Substituting (\ref{eq:a8}) into (\ref{eq:a7}) yields
$
m\dot s=-k y+\Delta m\dot z
$
with $\Delta m=\hat m-m$. The system can then be described by
\be
\begin{cases}
\Delta\dot x=\Delta z+ (1/\lambda_f)\dot y+ y\\
\Delta \dot z=\Delta \dot z\\
\Delta \ddot z=-\alpha_2\Delta \dot z-\alpha_1\Delta  z-\alpha_0\Delta x\\
m\dot s=-k y+\Delta m\dot z\\
\dot{\hat m}=-\gamma^\ast \dot z s
\end{cases}
\ee
with $\Delta x=x-x_d$ and $\Delta z=z-\dot x_d$.
The stability analysis can be completed by using similar procedures as in Sec. III.

\subsection{Consensus of Thrust-Propelled Vehicles}

Thrust-propelled vehicles represent a wide class of underactuated autonomous systems (see, e.g., \cite{Hua2009_TAC,Lee2012_AUT,Wang2016_TAC}); it is typically recognized that a TPV is an autonomous vehicle with its translational motion controlled by one thrust force along certain body-fixed direction
and its rotational motion fully actuated. We consider the consensus problem of $n$ TPVs, motivated by a well-known fact that a fundamental requirement in the formation problem is consensus (see, e.g., \cite{Lee2007_TAC}). The dynamics of the $i$-th TPV can be written as \cite{Lee2012_AUT,Pflimlin2007_IJC}
\begin{align}
\label{eq:44}
\begin{cases}
m_i \ddot{x}_i=-\sigma_i R_i e_3+m_i g e_3\\
\dot{R}_i=R_i S(\omega_i)
\end{cases}
\end{align}
where $\sigma_i\in {\mathcal R}$ is the thruster force, $m_i$ is the mass, $g$ denotes the gravitational acceleration, $e_3=\left[0, 0, 1\right]^T$, $x_i\in {\mathcal R}^3$ is the position vector of the TPV with respect to the inertial frame, $R_i\in \text{SO}(3)$ is the rotation matrix of the TPV with respect to the inertial frame, $\omega_i\in {\mathcal R}^3$ is the angular velocity of the TPV with respect to the inertial frame (expressed in the body-fixed frame), and the skew-symmetric form $S(\cdot)$ is defined as
$
S(b)=\begin{bmatrix}0 & -b_3 & b_2\\
b_3 & 0 & -b_1 \\
-b_2 & b_1 & 0 \end{bmatrix}
$
for a three-dimensional vector $b=\left[b_1,b_2, b_3\right]^T$.

 \subsubsection{Continuous Control}

 Using the dynamic feedback linearization in \cite{Wang2016_TAC}, we obtain that
\begin{align}
&m_i[\dddot x_i+(\alpha+\beta+\Sigma_{j\in\mathcal N_i(t)}w_{ij}(t))\ddot x_i]\nn\\
&-(\alpha+\beta+\Sigma_{j\in\mathcal N_i(t)}w_{ij}(t))m_i ge_3=u_i
\end{align}
where
\be
\label{eq:45}
u_i=-R_i\begin{bmatrix}\sigma_i \omega_i^{(2)}\\
-\sigma_i\omega_i^{(1)}\\
\dot \sigma_i+[\alpha+\beta+\Sigma_{j\in\mathcal N_i(t)}w_{ij}(t)]\sigma_i\end{bmatrix},
 \ee
$\omega_i^{(1)}$ and $\omega_i^{(2)}$ are the first and second components of $\omega_i$, respectively, and
 via (\ref{eq:45}), $\sigma_i$, $\omega_i^{(1)}$, and $\omega_i^{(2)}$ can be calculated by a typical procedure under the standard assumption that $\sigma_i\ne0$ after $u_i$ is specified.
 We design a control law for $u_i$ as
\begin{align}
u_i=&m_i \psi_i-[\alpha+\beta+\Sigma_{j\in\mathcal N_i(t)}w_{ij}(t)]m_i ge_3\\
\psi_i=&-\alpha\beta \dot x_i-\Sigma_{j\in\mathcal N_i(t)} w_{ij}(t)\nn\\
&\times[(\alpha+\beta)\dot x_i+\alpha\beta x_i-\alpha\beta x_j(t-T_{ij})],
\end{align}
which immediately yields
\begin{align}
\label{eq:46}
\dddot x_i=&-(\alpha+\beta)\ddot x_i-\alpha\beta \dot x_i\nn\\
&-\Sigma_{j\in\mathcal N_i(t)}w_{ij}(t)[\ddot x_i+(\alpha+\beta)\dot x_i\nn\\
&+\alpha\beta x_i-\alpha\beta x_j(t-T_{ij})].
\end{align}
Similar to the previous cases and with the same condition associated with the time-varying delay and interaction topology, by exploiting the Laplacian structure, we can directly obtain from (\ref{eq:46}) that $x_i-x_j\to 0$ and $\dot x_i\to 0$ as $t\to\infty$, $\forall i,j=1,\dots,n$.

\subsubsection{Differentiable Control}

To develop a differentiable control law, we employ a slightly different dynamic feedback linearization procedure (also based upon \cite{Wang2016_TAC}), namely
\begin{align}
\label{eq:47}
&m_i(\dddot x_i+\alpha^\ast\ddot x_i)-\alpha^\ast m_i ge_3=u_i^\ast
\end{align}
where $
u_i^\ast=-R_i[\sigma_i \omega_i^{(2)},
-\sigma_i\omega_i^{(1)},
\dot \sigma_i+\alpha^\ast\sigma_i]^T$.
Define a fourth-order dynamic system by adding two differentiators as
\begin{align}
\label{eq:49}
\dddot z_i=&-(\alpha+\beta+\gamma)\ddot z_i-(\alpha\beta+\alpha\gamma+\beta\gamma)\dot z_i\nn\\
&-\alpha\beta\gamma \dot x_i-\Sigma_{j\in\mathcal N_i(t)} w_{ij}(t)\nn\\
&\times[\ddot z_i+(\alpha+\beta+\gamma)\dot z_i+(\alpha\beta+\alpha\gamma+\beta\gamma)\dot x_i\nn\\
&+\alpha\beta\gamma x_i-\alpha\beta\gamma x_j(t-T_{ij})],
\end{align}
upon which we specify the control law as
\begin{align}
\label{eq:50}
u_i^\ast =m_i(\ddot z_i+\alpha^\ast \dot z_i)-k_i s_i^\ast -\alpha^\ast m_i ge_3.
\end{align}
where $k_i$ and $\gamma$ are positive design constants, and $
 s_i^\ast=\dot x_i-z_i
 $. This control law is differentiable due to (\ref{eq:49}).

 \emph{Theorem 4:} Suppose that there exist an infinite number of uniformly bounded intervals $[t_{\kappa_h},t_{\kappa_{h+1}})$, $h=1,2,\dots$ with $t_{\kappa_1}=t_0$ satisfying the property that the union of the interaction graphs in each interval contains a directed spanning tree. The controller given by  (\ref{eq:50}) with $z_i$ being given by (\ref{eq:49}) ensures the consensus of the $n$ TPVs, i.e., $x_i-x_j\to 0$ and $\dot x_i\to 0$ as $t\to\infty$, $\forall i,j=1,\dots,n$.

 \emph{Proof:}
 We obtain by substituting (\ref{eq:50}) into (\ref{eq:47}) that
\be
\label{eq:52}
m_i\ddot s_i^\ast=-\alpha^\ast m_i \dot s_i^\ast-k_i s_i^\ast.
\ee
Combining (\ref{eq:49}) and (\ref{eq:52}) yields
\begin{align}
\begin{cases}
\dfrac{d^4 x_i}{dt^4}=-(\alpha+\beta+\gamma)\ddot z_i-(\alpha\beta+\alpha\gamma+\beta\gamma)\dot z_i\\
\quad-\alpha\beta\gamma \dot x_i-\Sigma_{j\in\mathcal N_i(t)} w_{ij}(t)\\
\quad\times[\ddot z_i+(\alpha+\beta+\gamma)\dot z_i+(\alpha\beta+\alpha\gamma+\beta\gamma)\dot x_i\\
\quad+\alpha\beta\gamma x_i-\alpha\beta\gamma x_j(t-T_{ij})]+\dddot s_i^\ast\\
m_i\ddot s_i^\ast=-\alpha^\ast m_i \dot s_i^\ast-k_i s_i^\ast
\end{cases}
\end{align}
which is a dynamic-cascaded system with degree two. With the previous degree reduction transformation, we obtain that
\be
\label{eq:54}
\begin{cases}
\dot x_i=z_i+s_i^\ast\\
\dot z_i=\dot z_i\\
\ddot z_i=\ddot z_i\\
\dddot z_i=-(\alpha+\beta+\gamma)\ddot z_i-(\alpha\beta+\alpha\gamma+\beta\gamma)\dot z_i\\
\quad-\alpha\beta\gamma z_i-\Sigma_{j\in\mathcal N_i(t)} w_{ij}(t)\\
\quad\times[\ddot z_i+(\alpha+\beta+\gamma)\dot z_i+(\alpha\beta+\alpha\gamma+\beta\gamma)z_i\\
\quad+\alpha\beta\gamma x_i-\alpha\beta\gamma x_j(t-T_{ij})]\\
\quad-[\alpha\beta\gamma +(\alpha\beta+\alpha\gamma+\beta\gamma)\Sigma_{j\in\mathcal N_i(t)}w_{ij}(t)]s_i^\ast\\
m_i\ddot s_i^\ast=-\alpha^\ast m_i \dot s_i^\ast-k_i s_i^\ast
\end{cases}
\ee
which is a standard cascade system. The remaining proof can be completed using similar procedures as in Sec. III. \hfill{\small $\blacksquare$}

\subsubsection{Differentiable Adaptive Control}

In practice, the mass properties of the TPVs might be subjected to uncertainty due to, e.g., fuel consumption or load variation. This motivates the development of adaptive controllers (see, e.g., \cite{Wang2016_TAC}), and the adaptive controller in \cite{Wang2016_TAC}, however, relies on the measurement of translational acceleration. Here we develop an adaptive controller that does not rely on acceleration measurement. We propose the following adaptive controller
\begin{align}
\label{eq:57}
u_i^\ast=&\hat m_i(\ddot z_i+\alpha^\ast \dot z_i)-\alpha^\ast \hat m_i g e_3\\
\label{eq:58}
\dot {\hat m}_i=&-\gamma^\ast_i \left(\ddot z_i+\alpha^\ast \dot z_i-\alpha^\ast g e_3\right)^T \dot s_i^\ast
\end{align}
where $\hat m_i$ is the estimate of $m_i$ and $\gamma^\ast _i$ is a positive design constant. By exploiting the standard property of ``integral by parts'', we can directly obtain $\hat m_i$ without involving acceleration measurement, namely
\begin{align}
&\hat m_i
=\hat m_i(0)-\gamma_i^\ast\left.(\ddot z_i+\alpha^\ast \dot z_i-\alpha^\ast g e_3)^T \dot x_i\right.|_0^t\nn\\
&+\gamma_i^\ast\int_0^t [(\ddot z_i+\alpha^\ast \dot z_i-\alpha^\ast g e_3)^T\dot z_i +(\dddot z_i+\alpha^\ast \ddot z_i)^T \dot x_i]d\sigma.
\end{align}

\emph{Theorem 5:}  Suppose that there exist an infinite number of uniformly bounded intervals $[t_{\kappa_h},t_{\kappa_{h+1}})$, $h=1,2,\dots$ with $t_{\kappa_1}=t_0$ satisfying the property that the union of the interaction graphs in each interval contains a directed spanning tree. The adaptive  controller given by (\ref{eq:57}) and (\ref{eq:58}) with $z_i$ being given by (\ref{eq:49}) ensures the consensus of the $n$ TPVs, i.e., $x_i-x_j\to 0$ and $\dot x_i\to 0$ as $t\to\infty$, $\forall i,j=1,\dots,n$.

\emph{Proof:}
Substituting (\ref{eq:57}) into (\ref{eq:47}) yields
$
m_i\ddot s_i^\ast=-\alpha^\ast m_i \dot s_i^\ast+\Delta m_i(\ddot z_i+\alpha^\ast \dot z_i-\alpha^\ast g e_3)
$
where $\Delta m_i=\hat m_i-m_i$. As in the typical practice, we consider the Lyapunov-like function candidate
$
V_i^\ast=(m_i/2)\dot s_i^{\ast T} \dot s_i^\ast+[1/(2\gamma_i^\ast)](\Delta m_i)^2
$
and its derivative along the trajectories of the system can be written as
$\dot V_i^\ast=-\alpha^\ast m_i\dot s_i^{\ast T}\dot s_i^\ast\le 0
$. This immediately implies that $\dot s_i^\ast \in\mathcal L_2\cap \mathcal L_\infty$ and $\hat m_i\in\mathcal L_\infty$.
  We can obtain the following cascaded system by a degree reduction transformation that differs from the one used for deriving (\ref{eq:54})
\be
\begin{cases}
\dot x_i=\dot x_i\\
\ddot x_i=\dot z_i+\dot s_i^\ast\\
\ddot z_i=\ddot z_i\\
\dddot z_i=-(\alpha+\beta+\gamma)\ddot z_i-(\alpha\beta+\alpha\gamma+\beta\gamma)\dot z_i\\
\quad-\alpha\beta\gamma \dot x_i-\Sigma_{j\in\mathcal N_i(t)} w_{ij}(t)[\ddot z_i+(\alpha+\beta+\gamma)\dot z_i\\
 \quad+(\alpha\beta+\alpha\gamma+\beta\gamma)\dot x_i+\alpha\beta\gamma x_i-\alpha\beta\gamma x_j(t-T_{ij})]\\
m_i\ddot s_i^\ast=-\alpha^\ast m_i \dot s_i^\ast+\Delta m_i(\ddot z_i+\alpha^\ast \dot z_i-\alpha^\ast g e_3).
\end{cases}
\ee
The remaining proof can be completed with similar procedures as in Sec. III. \hfill {\small $\blacksquare$}

\emph{Remark 8:} The flexibility of forwardstepping leads us to derive an adaptive controller for networked TPVs without involving acceleration measurement, in contrast to the one in \cite{Wang2016_TAC}. The results in, e.g., \cite{Abdessameud2011_Aut,Lee2012_AUT}, based on the backstepping approach, are either confined to a fixed topology or involve the acceleration measurement (for instance, in \cite{Lee2012_AUT}, the acceleration measurement is actually required, which, though, can be obtained using the system model; this assumption of the exact model, however, might be restrictive). There are some results that do not rely on the backstepping approach, for instance, \cite{Yang2010_ICRA,Dong2010_IETCA,Borhaug2011_TCST}, which focus on the formation problem in SE(2) without considering parametric uncertainty and without handling time-varying delays and/or switching topologies.

\subsection{Trajectory Tracking Control of a Single Mechanical System}

In this subsection, we demonstrate the application of forwardstepping in trajectory tracking control of a single mechanical system (robotic system and spacecraft) without velocity measurement. As discussed in Sec. IV-B, for such physical systems, the reference dynamics can be chosen based on controlled physical dynamics with nonlinearity reduction.


\subsubsection{Task-Space Adaptive Control of Robotic Systems Without Task-Space Velocity Measurement}

Consider the task-space control problem of a robotic system with uncertain kinematics and dynamics, and the mapping from joint space to task space can be written as \cite{Craig2005_Book,Spong2006_Book}
\be
\label{eq:71}
x=f(q)
\ee
where $q\in \mathcal R^m$ is the joint position, $x\in\mathcal R^m$ is the task-space position, and $f: \mathcal R^m\to\mathcal R^m$ is a nonlinear mapping. Differentiating (\ref{eq:71}) with respect to time yields \cite{Craig2005_Book,Spong2006_Book}
\be
\label{eq:72}
\dot x=J(q)\dot q
\ee
where $J(q)\in \mathcal R^{m\times m}$ is the Jacobian matrix. The dynamics of the robotic system can be written as \cite{Slotine1991_Book,Spong2006_Book}
\be
\label{eq:73}
M(q)\ddot q+C(q,\dot q)\dot q+g(q)=\tau
\ee
where $M(q)\in \mathcal R^{m\times m}$ is the inertia matrix, $C(q,\dot q)\in\mathcal R^{m\times m}$ is the Coriolis and centrifugal matrix, $g(q)\in \mathcal R^m$ is the gravitational torque, and $\tau\in \mathcal R^m$ is the exerted joint torque.

\emph{Property 4 (\cite{Cheah2006_IJRR}):} The kinematics (\ref{eq:72}) depend linearly on an unknown constant kinematic parameter vector $\theta$, which yields
\be
\label{eq:74}
J(q)\xi^\ast=Z(q,\xi^\ast)\theta
\ee
where $\xi^\ast\in \mathcal R^m$ is a vector and $Z(q,\xi^\ast)$ is the kinematic regressor matrix.

\textit{Property 5 (\cite{Slotine1991_Book,Spong2006_Book}):} The inertia matrix $M (q )$ is symmetric and uniformly positive
definite.

\textit{Property 6 (\cite{Slotine1991_Book,Spong2006_Book}): }The Coriolis and centrifugal matrix $C(q,\dot q)$ can be
appropriately determined such that $\dot {M}(q) - 2C(q,\dot q)$ is skew-symmetric.

\textit{Property 7 (\cite{Slotine1991_Book,Spong2006_Book}): }The dynamics (\ref{eq:73}) depend linearly on an unknown
constant dynamic parameter vector $\vartheta$, which leads to
\begin{equation}
\label{eq:75}
M \left( q  \right)\dot \zeta ^\ast + C \left( q ,\dot {q} \right)\zeta ^\ast+ g \left( {q } \right) = Y ( q ,\dot {q}
,\zeta^\ast,\dot \zeta^\ast )\vartheta
\end{equation}
where $\zeta^\ast \in \mathcal R^m$ is a differentiable vector, $\dot {\zeta }^\ast$ is the derivative of $\zeta^\ast $, and $Y( q ,\dot {q} ,\zeta^\ast ,\dot {\zeta }^\ast )$ is the dynamic
regressor matrix.

Let $x_d\in\mathcal R^m$ denote the desired trajectory in the task space and assume that $x_d$, $\dot x_d$, and $\ddot x_d$ are bounded. 
 We introduce a nonlinear second-order dynamic system as
\be
\label{eq:76}
\dot z=\dot z_r-\alpha(z-z_r)-\hat J^T(q)K^\ast\Delta x
\ee
where $\Delta x=x-x_d$, $\hat J(q)$ is the estimate of $J(q)$ which is derived by replacing $\theta$ in $J(q)$ with its estimate $\hat\theta$, $K^\ast$ is a symmetric positive definite matrix, $\alpha$ is a positive design constant, and
\begin{align}
\label{eq:77}
z_r=&\hat J^{-1}(q)\dot x_d\\
\label{eq:78}
\dot z_r=&\hat J^{-1}(q)\left[\ddot x_d-\dot{\hat J}(q)z_r\right].
\end{align}
Define
\be
\label{eq:79}
s=\dot q-z.
\ee
The adaptive controller is given as
\begin{align}
\label{eq:80}
\tau=&-K s-\kappa \hat J^T(q)K^\ast\Delta x+Y(q,\dot q,z,\dot z)\hat\vartheta\\
\label{eq:81}
\dot{\hat \vartheta}=&-\Gamma Y^T(q,\dot q,z,\dot z)s\\
\label{eq:82}
\dot{\hat \theta}=&\Lambda Z^T(q,\dot q)K^\ast\Delta x
\end{align}
where $K$, $\Gamma$, and $\Lambda$ are symmetric positive definite matrices, $\kappa$ is a positive design constant, and $\hat\vartheta$ is the estimate of $\vartheta$.

\emph{Theorem 6:} The adaptive controller given by (\ref{eq:80}), (\ref{eq:81}), and (\ref{eq:82}) with $z$ being given by (\ref{eq:76}) for the robotic system given by (\ref{eq:72}) and (\ref{eq:73}) ensures the convergence of the task-space tracking errors, i.e., $\Delta x\to 0$ and $\Delta \dot x\to 0$ as $t\to\infty$.

\emph{Proof: } The dynamics of the system can be described by a dynamic-cascaded system
\be
\label{eq:a10}
\begin{cases}
\ddot q-\dot z_r=-\alpha(z-z_r)-\hat J^T(q)K^\ast\Delta x+\dot s\\
\dot{\hat \theta}=\Lambda Z^T(q,\dot q)K^\ast\Delta x\\
M(q)\dot s+C(q,\dot q)s\\
=-Ks-\kappa \hat J^T(q)K^\ast\Delta x +Y(q,\dot q,z,\dot z)\Delta \vartheta\\
\dot{\hat \vartheta}=-\Gamma Y^T(q,\dot q,z_r,\dot z_r)s
\end{cases}
\ee
where $\Delta \vartheta=\hat\vartheta-\vartheta$. Using degree reduction, we obtain that
 \be
\begin{cases}
\frac{d}{dt}(\dot q-z_r-s)=-\alpha(\dot q-z_r-s)-\hat J^T(q)K^\ast\Delta x\\
\dot{\hat \theta}=\Lambda Z^T(q,\dot q)K^\ast\Delta x\\
M(q)\dot s+C(q,\dot q)s\\
=-Ks-\kappa \hat J^T(q)K^\ast\Delta x +Y(q,\dot q,z,\dot z)\Delta \vartheta\\
\dot{\hat \vartheta}=-\Gamma Y^T(q,\dot q,z_r,\dot z_r)s.
\end{cases}
\ee
Consider the Lyapunov function candidate $V=({\kappa}/{2})[(z-z_r)^T(z-z_r)+\Delta x^T K^\ast\Delta x+\Delta \theta^T \Lambda^{-1}\Delta \theta]
+(1/2)s^T M(q)s+(1/2)\Delta\vartheta^T \Gamma^{-1}\Delta \vartheta
$
with $\Delta \theta=\hat\theta-\theta$, and the derivative of $V$ along the trajectories of the system can be written as (using Property 6)
$
\dot V=-s^T Ks-\kappa \alpha(z-z_r)^T (z-z_r)\le 0.
$
This implies that $s\in{\mathcal L}_2\cap{\mathcal L}_\infty$, $z-z_r\in{\mathcal L}_2\cap{\mathcal L}_\infty$, $\Delta x\in{\mathcal L}_\infty$, $\hat\vartheta\in{\mathcal L}_\infty$, and $\hat\theta\in{\mathcal L}_\infty$. From (\ref{eq:77}), we know that $z_r\in{\mathcal L}_\infty$ if $\hat J(q)$ is nonsingular, and thus $z\in{\mathcal L}_\infty$. This directly yields the result that $\dot q=s+z\in{\mathcal L}_\infty$. From (\ref{eq:82}), we obtain that $\dot{\hat\theta}\in{\mathcal L}_\infty$ and therefore $\dot{\hat J}(q)$ is bounded, which gives rise to the conclusion that $\dot z_r\in{\mathcal L}_\infty$ from (\ref{eq:78}). From (\ref{eq:76}), we can then directly obtain that $\dot z\in{\mathcal L}_\infty$. Hence, $z-z_r$ is uniformly continuous and in accordance with the properties of square-integrable and uniformly continuous functions \cite[p.~232]{Desoer1975_Book}, we obtain that $z-z_r\to 0$ as $t\to\infty$. From (\ref{eq:72}), we obtain that $\dot x\in{\mathcal L_\infty}$ and consequently $\Delta\dot x\in{\mathcal L}_\infty$. Therefore $\frac{d}{dt}[\hat J^T(q)K^\ast\Delta x]$ is bounded, implying that $\hat J^T(q)K^\ast\Delta x$ is uniformly continuous. We rewrite (\ref{eq:76}) as
$
 \dot z-\dot z_r=-\alpha(z-z_r)- \hat J^T(q)K^\ast\Delta x,
 $
 upon which we obtain that $\dot z-\dot z_r$ is uniformly continuous since the right side of the above equation is uniformly continuous. The application of Barbalat's lemma \cite{Slotine1991_Book} yields the result that $\dot z-\dot z_r\to 0$ as $t\to\infty$. As a consequence, $\hat J^T(q)K^\ast\Delta x\to 0$ as $t\to\infty$, which further implies that $\Delta x \to 0$ as $t\to\infty$ if $\hat J(q)$ is nonsingular.
 From the third subsystem of (\ref{eq:a10}) and using Property 5, we obtain that $\dot s\in{\mathcal L}_\infty$ and this directly leads to the result that $\ddot q\in{\mathcal L}_\infty$. Based on the differentiation of (\ref{eq:72}), we obtain that $\ddot x\in{\mathcal L}_\infty$ and hence $\Delta \ddot x\in{\mathcal L}_\infty$. This means that $\Delta \dot x$ is uniformly continuous. According to Barbalat's lemma \cite{Slotine1991_Book}, we obtain that $\Delta \dot x\to 0$ as $t\to\infty$. \hfill {\small $\blacksquare$}

\emph{Remark 9:} The adaptive tracking controller derived using forwardstepping extends the adaptive regulator in \cite{Wang2020_TAC} from an explicit dynamic-cascaded perspective. It does not involve task-space velocity measurement, in contrast with \cite{Cheah2006_IJRR,Liu2006_AUT,Wang2015_CCC} which either give rise to the overparametrization problem or require an observer. The avoidance of task-space velocity measurement is achieved by designing a second-order dynamic system (with respect to $q$) to yield dynamic feedback [i.e., using $z$ which is dynamically generated by (\ref{eq:76})]. The choice of the reference dynamics (\ref{eq:76}) is motivated in part by the controlled dynamics of robot manipulators in \cite{Takegaki1981_ASME} and the task-space regulator in \cite{Wang2020_TAC}. The forwardstepping design differs from most results in the literature that rely on the differential operation of joint reference velocity (e.g., \cite{Slotine1987_IJRR,Wang2017_TAC}).

\subsubsection{Global Attitude Tracking Control of Spacecraft Without Angular Velocity Measurement}

Consider a spacecraft actuated by reaction wheels, and the attitude dynamics of the spacecraft can be written as \cite{Slotine1990_TAC}
\be
\label{eq:94}
M\dot\omega-S(h)\omega=\tau
\ee
where $M\in {\mathcal R}^{3\times 3}$ is the inertia matrix, $h\in {\mathcal R}^3$ is the total angular momentum of the system expressed in the spacecraft frame which can be written as $h=R^T h_I$ with $h_I\in {\mathcal R}^3$ denoting the constant total angular momentum in the inertial frame and $R\in {\rm SO}(3)$ the rotation matrix of the spacecraft with respect to the inertial frame, $\omega\in \mathcal R^3$ is the angular velocity of the spacecraft expressed in the spacecraft frame, and $\tau\in {\mathcal R}^3$ is the exerted torque by the reaction wheels. Denote by $q_v\in\mathcal R^3$ and $q_o\in\mathcal R$ the vector and scalar parts of the Euler parameter vector associated with $R$, respectively, and the attitude kinematics can then be written as \cite{Egeland1994_TAC}
\begin{align}
\dot q_v=&\frac{1}{2}[q_o I_3+S(q_v)]\omega\\
\dot q_o=&-\frac{1}{2}q_v^T \omega.
\end{align}

Denote by $R_d\in{\rm SO}(3)$ the desired rotation matrix with $\dot R_d$ and $\ddot R_d$ being bounded, and our control objective is to achieve the convergence of the attitude tracking error without angular velocity measurement, i.e., $R\to R_d$ as $t\to\infty$. We introduce a nonlinear second-order dynamic system as
\be
\label{eq:97}
\dot z=R^T\dot  \omega_d^{I}-S(z)R^T \omega_d^I-\alpha_2 (z-\omega_d)-\alpha_1 \Delta q_v
\ee
where $\omega_d^I$ is the desired angular velocity in the inertial frame defined as (see, e.g., \cite{Egeland1994_TAC})
$
S(\omega_d^I)=\dot R_d R_d^T
$, $
\omega_d=R^T \omega_d^I$ is the desired angular velocity in the spacecraft frame, $\Delta q_v$ is the vector part of the Euler parameter vector (with $\Delta q_o$ denoting the scalar part) associated with the error rotation matrix $R_d^T R$, and $\alpha_1$ and $\alpha_2$ are positive design constants.

We propose the following control law
\be
\label{eq:99}
\tau=M\dot z-S(h)z- D^T(\Delta q^\ast)K \dot y
\ee
where the vector $y$ is given as
\begin{align}
\label{eq:100}
\dot y+\Lambda_f y=&K \Delta q^\ast,
\end{align}
$
D(\Delta q^\ast)=(1/2)\begin{bmatrix}\Delta q_o^\ast I_3+S(\Delta q_v^\ast)\\
-\Delta q_v^{\ast T}\end{bmatrix}
$, $K$ and $\Lambda_f$ are symmetric positive definite matrices, and $\Delta q^\ast=[\Delta q_v^{\ast T},\Delta q_o^\ast]^T$ is obtained by the Euler parameter multiplication (see, e.g., \cite{Ickes1970_AIAAJ})
\begin{align}
\Delta q_v^\ast=&q_{o,z} q_v-q_o q_{v,z}+S(q_v)q_{v,z}\\
\Delta q_o^\ast=&q_{o}q_{o,z}+q_{v}^T q_{v,z}
\end{align}
where $q_{v,z}\in \mathcal R^3$ and $q_{o,z}\in\mathcal R$ are updated by the Euler-parameter-based kinematics
\begin{align}
\label{eq:104}
\dot q_{v,z}=&\frac{1}{2}[q_{o,z} I_3-S(q_{v,z})]R z\\
\label{eq:105}
\dot q_{o,z}=&-\frac{1}{2}q_{v,z}^T Rz.
\end{align}
It can be shown by following the standard practice that
\be
\label{eq:106}
\Delta\dot q^\ast
={D(\Delta q^\ast)}(\omega-z).
\ee

\emph{Theorem 7:} The controller given by (\ref{eq:99})-(\ref{eq:105}) with $z$ being given by (\ref{eq:97}) ensures the convergence of the attitude tracking errors, i.e., $\Delta q_v\to 0$ and $\omega-\omega_d\to 0$ as $t\to\infty$.

\emph{Proof:} The dynamics of the system can be described by
\be
\label{eq:107}
\begin{cases}
\Delta\dot \omega=-\alpha_2\Delta \omega-\alpha _1\Delta q_v\\
\qquad\quad+(\dot \omega-\dot z)+\alpha_2(\omega-z)-S(\omega-z)\omega_d\\
M(\dot \omega-\dot z)=-S(h)(\omega-z)-D^T(\Delta q^\ast)K\dot y\\
\dot y+\Lambda_f y=K \Delta q^\ast
\end{cases}
\ee
where $\Delta \omega=\omega-\omega_d$. Consider the Lyapunov function candidate $V=(1/2)(\omega-z)^T M(\omega-z)+(1/2)\dot y^T \dot y$, and its derivative along the trajectories of the system can be written as $\dot V=-\dot y^T \Lambda_f \dot y\le 0$. This yields the conclusion that $\dot y\in{\mathcal L_2}\cap \mathcal L_\infty$ and $\omega-z\in\mathcal L_\infty$. From (\ref{eq:106}), we obtain that $\Delta \dot q^\ast\in\mathcal L_\infty$, yielding the result that $\ddot y\in\mathcal L_\infty$ from the differentiation of the third subsystem of (\ref{eq:107}), i.e.,
\be
\label{eq:108}
\ddot y+\Lambda_f \dot y=K\Delta \dot q^\ast.
\ee Hence, $\dot y$ and $\Delta q^\ast$ are uniformly continuous. From the properties of square-integrable and uniformly continuous functions \cite[p.~232]{Desoer1975_Book}, we obtain that $\dot y\to 0$ as $t\to\infty$. From the second subsystem of (\ref{eq:107}) and using the standard property that $M$ is positive definite, we obtain that $\dot \omega-\dot z\in\mathcal L_\infty$, and hence $\omega-z$ is uniformly continuous. Therefore, $\Delta \dot q^\ast$ is uniformly continuous based on (\ref{eq:106}), and this implies that $\ddot y$ is uniformly continuous from (\ref{eq:108}). Using Barbalat's lemma \cite{Slotine1991_Book}, we obtain that $\ddot y\to 0$ as $t\to\infty$, and hence that $\Delta \dot q^\ast\to 0$ as $t\to\infty$. By the standard practice (see, e.g., \cite{Egeland1994_TAC}), we obtain that $\omega-z\to 0$ as $t\to\infty$. From the second subsystem of (\ref{eq:107}), we obtain that $\dot \omega-\dot z\to 0$ as $t\to\infty$. Let us now analyze the first subsystem of (\ref{eq:107}). Due to the standard result that $\Delta \dot \omega=-\alpha_2\Delta \omega$ is exponentially stable, we obtain from the input-output properties of linear systems \cite[p.~59]{Desoer1975_Book} that $\Delta \omega\in\mathcal L_\infty$ and $\Delta \dot \omega\in\mathcal L_\infty$, which implies that the first subsystem of (\ref{eq:107}) is stable. At the extreme case $t\to\infty$, we have
$
\Delta \dot \omega=-\alpha_2\Delta\omega-\alpha_1\Delta q_v,
$
which can be shown to be asymptotically stable by the standard practice. In fact, consider the Lyapunov function candidate $V^\ast=(1/2)\Delta \omega^T \Delta \omega+2\alpha_1(1-\Delta q_o)$ and its derivative can be written as $\dot V^\ast=-\alpha_2\Delta \omega^T \Delta \omega\le 0$. Using the properties of square-integrable and uniformly continuous functions \cite[p.~232]{Desoer1975_Book} and Barbalat's lemma \cite{Slotine1991_Book}, we obtain that $\Delta \omega\to 0$ and $\Delta \dot \omega\to 0$ as $t\to\infty$, and hence $\Delta q_v\to 0$ as $t\to\infty$. \hfill {\small $\blacksquare$}

\emph{Remark 10:} Global attitude tracking control of spacecraft (or rigid bodies) without angular velocity measurement is achieved in \cite{Akella2001_SCL} based on either the modified Rodrigues parameter vector or Euler parameter vector (the case of the Euler parameter vector, as stated in \cite{Akella2001_SCL}, is essentially the same as that of the modified Rodrigues parameter vector). This approach relies closely on the properties of attitude dynamics given as $M\dot\omega+S(\omega)M\omega=\tau$ (see \cite{Akella2001_SCL}). In spacecraft engineering applications, reactions wheels are often used as actuators for attitude control, under which the spacecraft dynamics [i.e., equation (\ref{eq:94})] would differ from those considered in \cite{Akella2001_SCL}, yielding the consequence that the result in \cite{Akella2001_SCL} is not applicable. We solve this problem by designing a new class of dynamic feedback using the forwardstepping approach where the specified reference dynamic system (\ref{eq:97}) is motivated partly by the typical controlled dynamics of spacecraft with tracking control,  and this yields a globally stable attitude controller without using angular velocity measurement for spacecraft actuated by reaction wheels. The third term in (\ref{eq:99}) follows the passive-filter-based approach \cite{Berghuis1993_SCL,Lizarralde1996_TAC}. Our approach also differs from the locally stable approaches in, e.g., \cite{Caccavale1999_SCL,Costic2000_CDC}.

\subsection{Distributed Tracking of Networked Lagrangian Systems}

We here present solutions to the problem of distributed tracking of networked Lagrangian systems using forwardstepping. Let $q_0\in \mathcal R^m$ denote the generalized position of the (virtual) leader and assume that $\dot q_0$, $\ddot q_0$, and $\dddot q_0$ are bounded. We introduce the following definition of $z_i$ using nonlinear second-order reference dynamics
\begin{align}
\label{eq:110}
\dot z_i=&-\alpha \dot q_i-\Sigma_{j\in\mathcal N_i}w_{ij}(\xi_i-\xi_j)\nn\\
&-\gamma{\rm sgn}[\Sigma_{j\in\mathcal N_i}w_{ij}(\xi_i-\xi_j)]
\end{align}
with $\xi_0$ being defined as [similar to (\ref{eq:4})]
$
\xi_0=\dot q_0+\alpha q_0
$,
where $\gamma$ is a positive design constant to be determined in the sequel. Using the adaptive controller given by (\ref{eq:6}) and (\ref{eq:7}) with $z_i$ being given by (\ref{eq:110}) yields a dynamic-cascade system
\be
\label{eq:112}
\begin{cases}
\dot \xi_i=-\Sigma_{j\in\mathcal N_i}w_{ij}(\xi_i-\xi_j)-\gamma{\rm sgn}[\Sigma_{j\in\mathcal N_i}w_{ij}(\xi_i-\xi_j)]+\dot s_i\\
M_i(q_i)\dot s_i+C_i(q_i,\dot q_i)s_i=-K_i s_i+Y_i(q_i,\dot q_i,z_i,\dot z_i)\Delta \vartheta_i\\
\dot {\hat \vartheta}_i=-\Gamma_i Y_i^T(q_i,\dot q_i,z_i,\dot z_i)s_i.
\end{cases}
\ee

\emph{Theorem 8:} Suppose that $\gamma$ is chosen such that
$
\label{eq:113}
\gamma>\sup_t\{|\dot \xi_0(t)|_\infty\}
$
where $|\cdot|_\infty$ denotes the standard infinite norm, and that the interaction graph among the $n$ systems and the (virtual) leader (denoted by vertex 0) contains a directed spanning tree with the rooted vertex being $0$. Then, the adaptive controller given by (\ref{eq:6}) and (\ref{eq:7}) with $z_i$ being given by (\ref{eq:110}) ensures that $q_i-q_0\to 0$ and $\dot q_i-\dot q_0\to 0$ as $t\to\infty$, $\forall i=1,\dots,n$.

\emph{Proof:} We can transform (\ref{eq:112}) to the following system
\be
\label{eq:114}
\begin{cases}
\Delta \dot \xi_i=-\Sigma_{j\in\mathcal N_i}w_{ij}(\Delta\xi_i-\Delta\xi_j)\\
\qquad-\gamma{\rm sgn}[\Sigma_{j\in\mathcal N_i}w_{ij}(\Delta\xi_i-\Delta\xi_j)]-\dot \xi_0+\dot s_i\\
M_i(q_i)\dot s_i+C_i(q_i,\dot q_i)s_i=-K_i s_i+Y_i(q_i,\dot q_i,z_i,\dot z_i)\Delta \vartheta_i\\
\dot {\hat \vartheta}_i=-\Gamma_i Y_i^T(q_i,\dot q_i,z_i,\dot z_i)s_i
\end{cases}
\ee
where $\Delta \xi_i=\xi_i-\xi_0$, $i=1,\dots,n$ and $\Delta \xi_0=0$. With the same analysis as in the proof of Theorem 1, we obtain that $s_i\in\mathcal L_2\cap\mathcal L_\infty$ and $\hat\vartheta_i\in\mathcal L_\infty$, $\forall i$. {Using a degree reduction}, the first subsystem of (\ref{eq:114}) can be rewritten as
\begin{align}
\label{eq:115}
&\frac{d}{dt}(\Delta \xi_i-s_i)=-\Sigma_{j\in\mathcal N_i}w_{ij}[(\Delta\xi_i-s_i)-(\Delta\xi_j-s_j)]\nn\\
&-\gamma{\rm sgn}[\Sigma_{j\in\mathcal N_i}w_{ij}((\Delta\xi_i-s_i)-(\Delta\xi_j-s_j)+(s_i-s_j))]\nn\\
&-\Sigma_{j\in\mathcal N_i}w_{ij}(s_i-s_j)-\dot \xi_0,
\end{align}
where $s_0=0$, and since the linear system
\begin{align}
\frac{d}{dt}(\Delta \xi_i-s_i)=&-\Sigma_{j\in\mathcal N_i}w_{ij}[(\Delta\xi_i-s_i)-(\Delta\xi_j-s_j)],\nn\\
&\qquad \qquad\qquad\qquad i=1,\dots,n
\end{align}
is exponentially stable with $\Delta \xi_i-s_i$ as the state (see, e.g., \cite{Ren2008_Book}), we thus obtain that $\Delta \xi_i-s_i\in\mathcal L_\infty$ using the input-output properties of linear systems \cite[p.~59]{Desoer1975_Book}, $\forall i$. Hence $\Delta \xi_i\in\mathcal L_\infty$, and $\Sigma_{j\in\mathcal N_i}w_{ij}(\xi_i-\xi_j)\in\mathcal L_\infty$, $\forall i$.
From (\ref{eq:110}), we obtain
\begin{align}
\dot z_i=&-\alpha z_i-\Sigma_{j\in\mathcal N_i}w_{ij}(\xi_i-\xi_j)\nn\\
&-\gamma{\rm sgn}[\Sigma_{j\in\mathcal N_i}w_{ij}(\xi_i-\xi_j)]-\alpha s_i
,\end{align}
upon which we obtain that $z_i\in\mathcal L_\infty$ and $\dot z_i\in\mathcal L_\infty$ from the input-output properties of linear systems \cite[p.~59]{Desoer1975_Book}, $\forall i$. Hence $\dot q_i\in\mathcal L_\infty$, $\forall i$. From the second subsystem of (\ref{eq:114}), we obtain that $\dot s_i\in\mathcal L_\infty$ using Property 1, implying that $s_i$ is uniformly continuous, $\forall i$. Using the properties of square-integrable and uniformly continuous functions \cite[p.~232]{Desoer1975_Book}, we obtain that $s_i\to 0$ as $t\to\infty$, $\forall i$, under which equation (\ref{eq:115}) becomes
\begin{align}
\label{eq:118}
&\frac{d}{dt}(\Delta \xi_i-s_i)=-\Sigma_{j\in\mathcal N_i}w_{ij}[(\Delta\xi_i-s_i)-(\Delta\xi_j-s_j)]\nn\\
&-\gamma{\rm sgn}[\Sigma_{j\in\mathcal N_i}w_{ij}((\Delta\xi_i-s_i)-(\Delta\xi_j-s_j)]-\dot \xi_0,
\end{align}
 and under the condition associated with $\gamma$, we obtain from \cite{Cao2010_SCL} that $\Delta \xi_i-s_i\to 0$, and hence $\Delta \xi_i\to 0$ as $t\to\infty$, $\forall i$. From the standard input-output properties of linear systems, we have that $q_i-q_0\to 0$ and $\dot q_i-\dot q_0\to 0$ as $t\to\infty$, $\forall i$.    \hfill {\small $\blacksquare$}

Due to the involvement of the signum function in the above definition of $\dot z_i$, the control torques involve the chattering issue \cite{Slotine1991_Book}. For yielding differentiable control torques, we define nonlinear third-order reference dynamics as
\begin{align}
\label{eq:119}
\ddot z_i=&-\beta\ddot q_i-\alpha \dot q_i-\Sigma_{j\in\mathcal N_i}w_{ij}(\xi_i^\ast-\xi_j^\ast)\nn\\
&-\gamma{\rm sgn}[\Sigma_{j\in\mathcal N_i}w_{ij}(\xi_i^\ast-\xi_j^\ast)]
\end{align}
with $\xi_i^\ast=\ddot q_i+\beta \dot q_i+\alpha q_i$, $i=1,\dots,n$ and $\xi_0^\ast=\ddot q_0+\beta \dot q_0+\alpha q_0$.

\emph{Theorem 9:} Suppose that $\gamma$ is chosen such that
$
\gamma>\sup_t\{|\dot \xi_0^\ast(t)|_\infty\},
  $ and that the interaction graph among the $n$ systems and the (virtual) leader (denoted by vertex 0) contains a directed spanning tree with the rooted vertex being $0$. Then, the adaptive controller given by (\ref{eq:6}) and (\ref{eq:7}) with $z_i$ being given by (\ref{eq:119}) ensures that $q_i-q_0\to 0$ and $\dot q_i-\dot q_0\to 0$ as $t\to\infty$, $\forall i=1,\dots,n$.

The proof of Theorem 9 can be conducted by a similar procedure as in the proof of Theorem 8.

\emph{Remark 11:}
Distributed tracking is an important problem in the literature; the solutions in the case of linear systems appear in, e.g., \cite{Cao2010_SCL,Cao2012_TAC}, and for nonlinear uncertain systems, most results are based on a distributed observer (see, e.g., \cite{Mei2012_Aut,Meng2014_TRO}) or confined to undirected graphs (e.g., \cite{Ghapani2016_AUT}). The handling of time-varying reference also occurs in the context of distributed average tracking on undirected graphs (see, e.g., \cite{Chen2015_TAC,Ghapani2018_TAC}). To the best of our knowledge, the problem of distributed tracking using physical measurement only, either for linear systems with differentiable control or for nonlinear uncertain systems on directed graphs, remains open. {Our result gives solutions to this open problem using dynamic feedback and from a differential/dynamic-cascaded perspective}. In particular, our result depends on physical measurement only and does not rely on a distributed observer, and in addition the second solution yields differentiable control torques owing to the benefit of forwardstepping design. The adoption of integral of signum-function-based action is inspired by the results for control of systems with disturbances in, e.g., \cite{Xian2004_TAC,Patre2010_AUT}. Also note that the solution in \cite{Ghapani2016_AUT} for the swarm tracking problem can similarly be shaped to be differentiable using forwardstepping.

\emph{Remark 12:} In the differentiable adaptive controller given by (\ref{eq:6}), (\ref{eq:7}), and (\ref{eq:119}), the calculation of $\dot z_i$ involves the acceleration measurement. 
The avoidance of acceleration measurement can be achieved by redefining $z_i$ as
\begin{align}
\label{eq:122}
\dot z_i=&-\beta\dot q_i-\alpha q_i-\Sigma_{j\in\mathcal N_i}w_{ij}(\dot q_i-\dot q_j)\nn\\
&-\Sigma_{j\in\mathcal N_i}w_{ij}\int_0^t (\beta \dot q_i+\alpha q_i-\beta \dot q_j-\alpha q_j)d\sigma\nn\\
&{-\gamma\int_0^t {\rm sgn}[\Sigma_{j\in\mathcal N_i}w_{ij}(\xi_i^{\ast\ast}-\xi_j^{\ast\ast})]d\sigma}+c_i^{\ast\ast}
\end{align}
where $\xi_i^{\ast\ast}=\dot z_i+\beta \dot q_i+\alpha q_i$. This, on the other hand, would require the communication of the quantities $\dot z_i$, $i=1,\dots,n$ among the systems. However, the solution with $z_i$ given by (\ref{eq:122}) relies on dynamic feedback where $\dot z_i$ is strongly coupled to the physical quantities of the systems, and thus differs from the distributed-observer-based solutions in, e.g., \cite{Mei2012_Aut,Meng2014_TRO}.

\section{Conclusion}

In this paper, we start by investigating consensus problem for networked Lagrangian systems with arbitrary bounded time-varying delay and switching topology. Using a class of dynamic feedback that can yield arbitrary times differentiable reference velocities, we develop delay/topology-independent adaptive controllers with piecewise continuous or arbitrary times differentiable control torques, extending/unifying the existing results for consensus of Lagrangian systems. The condition required for realizing consensus among the systems is that the time-varying delay is piecewise uniformly continuous and uniformly bounded and the union of the interaction graphs in each of an infinite sequence of time intervals contains a directed spanning tree. Motivated by this starting example, we propose a new paradigm referred to as forwardstepping, which is shown to be a potential systematic tool for solving a wide class of control problems in other contexts, for instance, consensus of networked underactuated TPVs, trajectory control of a single mechanical system without velocity measurement (task-space velocity or angular velocity), and handling discontinuities of the signum-function-based control action in distributed tracking of networked uncertain Lagrangian systems. The forwardstepping approach basically relies on the integral action (and algebraic action) concerning the reference dynamics to generate the control input and typically yields differential/dynamic-cascaded closed-loop dynamics. The differential-cascaded closed-loop system yielded by forwardstepping shows a structure involving the linear and nonlinear integral action of the target dynamics over controlled dynamics around the input.


%







\bibliographystyle{IEEEtran}
\bibliography{..//Reference_list_Wang}

%
%
%

%








\end{document}